\documentclass[12pt]{article}
\pdfoutput=1
\usepackage[margin=1in]{geometry}
\usepackage{amsfonts,amssymb,epsfig,amsmath}
\usepackage{slashed}
\usepackage{color}
\usepackage{hyperref}

\renewcommand{\text}[1]{#1}

\newcommand{\be}{\begin{equation}}
\newcommand{\ee}{\end{equation}}
\newcommand{\ben}{\begin{displaymath}}
\newcommand{\een}{\end{displaymath}}
\newcommand{\bea}{\begin{eqnarray}}
\newcommand{\eea}{\end{eqnarray}}
\newcommand{\bean}{\begin{eqnarray*}}
\newcommand{\eean}{\end{eqnarray*}}
\newcommand{\nn}{\nonumber \\}
\newcommand{\ba}{\begin{array}}
\newcommand{\ea}{\end{array}}
\newcommand{\bi}{\begin{itemize}}
\newcommand{\ei}{\end{itemize}}

\def\b{\beta}
\def\g{\gamma}
\def\G{\Gamma}

\def\G{\Gamma}
\def\g{\gamma}
\def\e{\epsilon}
\def\s{\sigma}
\def\e{\epsilon}

\DeclareMathOperator{\vol}{vol}

\newcommand{\dd}{\mathrm{d}}
\newcommand{\DD}{\mathrm{D}}

\begin{document}


\begin{titlepage}

\vfill
\begin{flushright}
DMUS-MP-15-04 \\
YITP-SB-15-04
\end{flushright}

\vfill

\begin{center}
   \baselineskip=16pt
   {\Large \bf All supersymmetric solutions of 3D U(1)$^3$ gauged supergravity. }
   \vskip 2cm
     Eoin \'O Colg\'ain$^{a, b}$
       \vskip .6cm
             \begin{small}                 
                 \textit{$^a$ C.N.Yang Institute for Theoretical Physics, SUNY Stony Brook, NY 11794-3840, USA}
                 \vspace{3mm}
                 
                  \textit{$^b$ Department of Mathematics, University of Surrey, Guildford GU2 7XH, UK}
                 \vspace{3mm}

             \end{small}
\end{center}

\vfill \begin{center} \textbf{Abstract}\end{center} \begin{quote}
D3-branes wrapping constant curvature Riemann surfaces give rise to 2D $\mathcal{N} = (0,2)$ SCFTs, where the superconformal fixed-points are mapped to vacua of 3D $\mathcal{N} =2$ U(1)$^3$ gauged supergravity. In this work we determine the fermionic supersymmetry variations of the theory and present all supersymmetric solutions. For spacetimes with a timelike Killing vector, we identify new timelike warped AdS$_3$ (G\"odel) and timelike warped dS$_3$ fixed-points. We outline the construction of numerical solutions interpolating between fixed-points, demonstrate that these flows are driven by an irrelevant scalar operator in the SCFT and identify the inverse of the superpotential as a candidate $c$-function. We further classify all spacetimes with a null Killing vector, in the process producing loci in parameter space where null-warped AdS$_3$ vacua with Schr\"{o}dinger $z=2$ symmetry exist. We construct non-supersymmetric spacelike warped AdS$_3$ geometries based on D3-branes. 

 \end{quote} \vfill

\end{titlepage}



\section{Summary \& Outlook}
\setcounter{equation}{0}
Given a 4D gauge theory with $\mathcal{N}=1$ supersymmetry, 2D theories with $\mathcal{N} = (0,2)$ supersymmetry can be engineered by ``twisting" the theory, or in other words, coupling it to background gauge field, and reducing the theory on a Riemann surface. Through this step, it may be expected that the 2D theory inherits properties from the 4D parent. In fact, there are strong similarities; 2D dualities \cite{Gadde:2013lxa, Gadde:2014ppa,Kutasov:2013ffl,Kutasov:2014hha} bear a resemblance to 4D Seiberg duality \cite{Seiberg:1994pq}, and a 2D procedure to compute the exact central charge and R symmetry at superconformal fixed-points, $c$-extremization \cite{Benini:2012cz, Benini:2013cda} \footnote{The supergravity dual of $c$-extremization at the two-derivative level was discussed in \cite{Karndumri:2013iqa, Karndumri:2013dca}. $c$-extremization to include subleading terms appeared in \cite{Baggio:2014hua}. } is a 4D analogue of $a$-maximization \cite{Intriligator:2003jj}. More generally, 2D $\mathcal{N} = (0,2)$ theories merit study in their own right as they have applications to compactifications of heterotic string theory (see \cite{McOrist:2010ae} for a review). 

In this paper we will be specifically interested in twisted compactifications of $\mathcal{N}=4$ super-Yang-Mills \cite{Bershadsky:1995vm, Maldacena:2000mw} on a genus $\frak{g}$ Riemann surface of constant curvature $\kappa$, $\Sigma_{\frak{g}}$, giving rise to 2D $\mathcal{N} = (0,2)$ SCFTs in the low-energy limit. From the perspective of string theory, these theories correspond to D3-branes wrapping $\Sigma_{\frak{g}}$, where the spin connection of $\Sigma_{\frak{g}}$ is traded off against a background R symmetry gauge fields with constant twist parameters $a_I$ resulting in preserved supersymmetry provided
\be
\label{susy_cond}
a_1 + a_2 + a_3 = - \kappa. 
\ee
Over the last number of years, 2D SCFTs and their AdS$_3$ supergravity dual geometries have been studied in a host of papers \cite{Benini:2012cz, Benini:2013cda, oai:arXiv.org:1108.1213, Donos:2011pn} (see also earlier \cite{Naka:2002jz, oai:arXiv.org:hep-th/0303211}). As string theory can be neatly truncated to 3D $\mathcal{N}=2$ U(1)$^3$ gauged supergravity \cite{Karndumri:2013iqa, Karndumri:2013dca}, 3D supergravity provides an overarching description of these vacua and their supersymmetric deformations. Recent generalisations of this construction include twisted compactifications from less supersymmetric $\mathcal{N} =1$ 4D theories \cite{Bobev:2014jva} and extensions to Riemann surfaces with boundaries \cite{Nagasaki:2014xya}. 

In this work, we consider the dimensional reduction of the accompanying fermionic supersymmetry variations for the purely bosonic consistent truncation presented in  ref. \cite{Karndumri:2013iqa, Karndumri:2013dca}. Traditionally, such reductions are often overlooked, since it is usually easier to reconstruct the fermionic sector from the bosonic sector and a knowledge of the supergravity.  In the context of 3D gauged supergravity, this approach was adopted in \cite{Colgain:2010rg}. That being said, prominent examples of full reductions exist \cite{deWit:1986iy,Nastase:1999kf, Nastase:1999cb} and one usually gains added insights by reducing the fermionic supersymmetry variations  \cite{Buchel:2006gb, Gauntlett:2006ai, Bah:2010cu,Bah:2010yt, Szepietowski:2012tb}. In section \ref{sec:susy}, we confirm that the expected superpotential of 3D U(1)$^3$ gauged supergravity also falls out of the fermionic supersymmetry variations, thus confirming the identity of the lower-dimensional theory. Setting these variations to zero, we identify the Killing spinor equations of the gauged supergravity. 

With the Killing spinor equations in hand, it is feasible to extract all the supersymmetric solutions in Lorentzian signature. To do so, one makes use of powerful Killling spinor techniques to recast the supersymmetry conditions in the natural language of differential geometry. Following the pioneering work of Tod \cite{Tod:1983pm} in 4D, this approach has been well-honed in 5D, where it has led to a host of beautiful results, including the discovery of G\"{o}del universe with enhanced supersymmetry \cite{Gauntlett:2002nw}, a supersymmetric black ring \cite{Elvang:2004rt}, concentric black rings \cite{Gauntlett:2004wh, Gauntlett:2004qy} and AdS$_5$ black holes \cite{Gutowski:2004ez, Gutowski:2004yv}. Here, we provide an analogous treatment for 3D U(1)$^3$ gauged supergravity. The same approach has recently been applied to classify solutions to 3D maximal and half-maximal supergravity \cite{Deger:2010rb, deBoer:2014iba, Deger:2015tra}. 

For spacetimes admitting a timelike Killing vector, we show that supersymmetric solutions are completely determined by a set of equations, one for each scalar, making three in total, and a Liouville-type equation for a Riemann surface. The remaining equations are implied by supersymmetry. Away from the AdS$_3$ fixed-point, the solutions all preserve half of the supersymmetries. Interestingly, in addition to the supersymmetric AdS$_3$ solution, new fixed-points exist, which are not critical points of the superpotential, yet the scalars are constant. As reported in \cite{Colgain:2015ela}, these solutions only exist at points in parameter space where the internal Riemann surface is hyperbolic, $\Sigma_{\frak{g}} =$ H$^2/\Gamma$ $(\frak{g} > 1)$, where the quotient is performed with respect to a subgroup $\Gamma$ of SL(2, $\mathbb{R})$ \footnote{For the 10D geometry to be well-defined we require $2 a_{I} (\frak{g}-1) \in \mathbb{Z}$. }. In one region, the fixed-points correspond to 3D G\"{o}del universes \cite{Godel, Reboucas:1982hn} \footnote{A number of supersymmetric embeddings of 3D G\"{o}del in string theory exist \cite{Israel:2003cx,Compere:2008cw,Levi:2009az}.}, while in another they are topologically $\mathbb{R} \times$S$^2$.  All new fixed-points exhibit closed-timelike curves (CTCs). The fixed-points are closely related to 5D solutions with a product base H$^2 \times$H$^2$ and S$^2 \times$H$^2$ discussed in \cite{Gauntlett:2003fk}, although the only overlap occurs at an isolated point in parameter space, where there are no G\"{o}del solutions. In addition, we remark that when $\frak{g} \leq 1$, the only explicit timelike solution we are aware of is the AdS$_3$ vacuum, making it of interest to find others. It would also be interesting to find black hole solutions \footnote{In 5D the natural black hole ansatz involves a squashed S$^3$. As we have assumed that the 5D spacetime is a warped product $M_{1,2} \times \Sigma_{\frak{g}}$, our 5D spacetimes fail to cover the natural black hole ansatz.}. 

Our G\"{o}del solutions provide families of 3D $\mathcal{N}=2$ gauged supergravity examples with a 5D uplift. It is expected that they generalise solutions originally reported in \cite{Banados:2007sq} in a purely 3D context.  Furthermore, like \cite{Banados:2007sq}, our embedding appears to preclude so-called G\"{o}del black holes \cite{Banados:2005da}, making it of interest to find such gauged supergravity embeddings \footnote{The analogous analysis in 7D gauged supergravity where the lower-dimensional 3D theory has an [SU(1,1)/U(1)]$^4$ scalar manifold appears in \cite{eoin}.}. The new fixed-points may be analytically continued in a number of interesting ways. Analytically continuing $\mathbb{R} \times$S$^2$ fixed-points, we get squashed three-spheres, or Berger spheres. We can also imagine analytically continuing $\mathbb{R} \times$H$^2$ to give spacelike warped AdS$_3$ with a U(1) fibre over an AdS$_2$ base, S$^1 \times$AdS$_2$. Within the context of our consistent truncation, the price one pays is that one has either to consider a complexification of the Chern-Simons coefficients, i. e. a complex theory, or one finds that the solution is supported by a  complex flux. In the appendix we relax supersymmetry and identify spacelike warped AdS$_3$ vacua, which embed in string theory, thus providing an alternative construction of these geometries that does not rely on T-duality, e. g. \cite{Orlando:2010ay}. 

In section \ref{sec:SUSYflow}, we show that one can find numerical interpolating solutions between fixed-points, which are driven by irrelevant scalar operators in the SCFT. It is known that in the vicinity of a superconformal fixed-point, the inverse of the superpotential corresponds to Zamolodchikov's $c$-function \cite{Zamolodchikov:1986gt}, which gets extremised in the process of $c$-extremization \cite{Karndumri:2013iqa}. It is easy to see that in flows from AdS$_3$ to G\"{o}del this same function decreases, however if one directly inputs any of our explicit G\"{o}del solutions into the central charge of ref. \cite{Compere:2007in}, one recovers the AdS$_3$ value for the central charge \cite{Brown:1986nw}. The likely resolution of this apparent contradiction is that the results of \cite{Compere:2007in} should be revisited and generalised to our setting. We have overlooked flows to $\mathbb{R} \times$S$^2$ fixed-points in this discussion as they exhibit topology change and it is unlikely that a monotonically decreasing function exists. 

Finally, we classify spacetimes with a null Killing vector, which cover well-known flows from AdS$_5$ to AdS$_3$ \cite{Maldacena:2000mw}. The analysis presents a refinement of the more general 5D results of ref. \cite{Gutowski:2005id} tailored to direct-products of a Riemann surface with a 3D spacetime. As an application, we identify loci in parameter space where null-warped AdS$_3$, or Schr\"{o}dinger solutions with dynamical exponent $z=2$ \cite{Son:2008ye,Balasubramanian:2008dm}, exist. The solutions preserve only a single supersymmetry, and in contrast to deformations based on D1-D5 \cite{Jeong:2014iva}, there is not enough preserved supersymmetry to identify a corresponding Schr\"odinger superalgebra  \footnote{On dimensionality grounds, we expect these theories to be dual to quantum mechanical systems. Examples with $\mathcal{N}=2$ Schr\"{o}dinger symmetry are known to exist \cite{Galajinsky:2009dw}.}. It is an interesting feature of the solutions that null-warped AdS$_3$ vacua appear precisely along the loci where G\"{o}del fixed-points become AdS$_3$. It is expected that these solutions can be traced to a subsector of $\mathcal{N}=4$ super-Yang-Mills deformed by an irrelevant operator \cite{Guica:2010sw}. 
 
The structure of the paper is as follows. Following a lightning review of 3D U(1)$^3$ gauged supergravity in the next section, in sections \ref{sec:susy} and \ref{sec:integrability}, we dimensionally reduce the fermionic supersymmetry variations from 5D and show through integrability that the resulting Killing spinor equations are consistent with the equations of motion (EOMs) of the bosonic sector. In section \ref{sec:classify} we present the results of our classifications, while in section \ref{sec:SUSYflow}, we construct numerical flows interpolating between sample timelike fixed-points. In section \ref{sec:5Dconn} we illustrate how the solutions to 3D U(1)$^3$ gauged supergravity embed in a well-known classification of 5D U(1)$^3$ gauged supergravity \cite{Gutowski:2004yv}. Our conventions, further details of the EOMs and a construction of spacelike warped AdS$_3$ can be found in the appendix.  

\subsection*{Review of 3D theory} 
This work concerns a consistent truncation of string theory to a 3D supergravity theory, which we refer to as 3D U(1)$^3$ gauged supergravity. The truncation of the bosonic sector was already featured in \cite{Karndumri:2013iqa, Karndumri:2013dca}, where the lower-dimensional theory was demonstrated to be consistent with the structure of 3D $\mathcal{N} =2$ gauged supergravity \cite{deWit:2003ja}, a theory possessing a K\"{a}hler scalar manifold, and thus an even number of scalars. The theory may be uplifted on a (constant curvature) genus $\frak{g}$ Riemann surface $\Sigma_{\frak{g}}$ to 5D U(1)$^3$ gauged supergravity, a well-known consistent truncation of string theory on S$^5$ \cite{Cvetic:1999xp} \footnote{As 5D U(1)$^3$ gauged supergravity can be truncated to minimal gauged supergravity, for the special choice of equal $a_I$, the 3D theory can be embedded in a universal way \cite{Gauntlett:2006ai, Gauntlett:2007ma}. }. The bosonic sector of the same theory also arises as a consistent truncation of 11D supergravity on three disks \cite{Colgain:2014pha}, but the embedding breaks supersymmetry \footnote{Interestingly, once the gauge fields are truncated out, the embedding is also Ricci-flat}.

The dimensionally reduced 3D theory may be expressed as \cite{Karndumri:2013iqa, Karndumri:2013dca} \bea
\label{Einsteinact}
\mathcal{L}_3 &=&  R *_3 \mathbf{1} - \frac{1}{2} \sum_{I=1}^3 \left[ \dd W_I \wedge *_3 \dd W_I + e^{2 W_I} G^I
\wedge *_3 G^I \right]   \nn &+&
4 g^2 \left[
e^{-W_1-W_3} + e^{-W_2  -W_3} + e^{-W_1-W_2} \right]  +  2 \kappa
e^{-W_1 -W_2 -W_3} \nn &-& \frac{1}{2} \left[ a_1^2 \,
e^{-2(W_2+W_3)} + a_2^2 \, e^{-2(W_1+W_3)} + a_3^2 \,
e^{-2(W_1+W_2)} \right] \nn &-& a_1 B^2 \wedge \dd B^3 - a_2 B^3 \wedge \dd B^1
- a_3 B^1 \wedge \dd B^2,  
\eea
where $g$ is a coupling constant, inherited from the 5D theory, which we henceforth normalise to unity, $\kappa$ is the constant curvature of $\Sigma_{\frak{g}}$, the internal Riemann surface appearing in the reduction from 5D, and $a_I$, $I=1, 2, 3$ correspond to twist parameters in the dual field theory \cite{Bershadsky:1995vm, Vafa:1994tf}. The field content of the theory comprises three scalars, $W_I$,  and three gauge fields, $B^I$, with field strengths, $G^I = \dd B^I$. In terms of the breathing mode of $\Sigma_{\frak{g}}$, $C$, and the scalars of the original 5D theory, $W_I$ may be written as 
\bea
\label{scalars3D5D}
W_1 &=& 2 C + \tfrac{1}{\sqrt{6}} \varphi_1 + \tfrac{1}{\sqrt{2}} \varphi_2, ~~W_2 = 2 C + \tfrac{1}{\sqrt{6}} \varphi_1 - \tfrac{1}{\sqrt{2}} \varphi_2, ~~W_3 = 2 C - \tfrac{2}{\sqrt{6}} \varphi_1. 
\eea
\textit{A priori}, the 3D action does not correspond to a supergravity, unless $\kappa$, the curvature of $\Sigma_{\frak{g}}$ satisfies the constraint (\ref{susy_cond}). In this case, one can introduce a real superpotential, $T$ \footnote{The potential and superpotential for $\frak{g} \neq 1$ originally appeared in \cite{oai:arXiv.org:hep-th/0303211}.}, 
\be
\label{T}
T = \sum_{I=1}^3 (\frac{1}{2} e^{-W_I} - \frac{1}{4} e^{K} a_I e^{W_I} )
\ee
where $K$ is the K\"{a}hler potential $K = - (W_1 + W_2 + W_3)$ of the 3D gauged supergravity and rewrite the action in the canonical form of a non-linear sigma model coupled to supergravity \cite{Karndumri:2013iqa}
\bea
\mathcal{L}_3 &=& R *_3 \mathbf{1} - g_{I \bar{J}} \DD z_{I} \wedge *_3 \DD \bar{z}_J + \left( 8  T^2 - 8 g^{I \bar{J}} \partial_I T \partial_{\bar{J}} T \right) *_3 \mathbf{1}  \nn &-& a_1 B^2 \wedge \dd B^3 - a_2 B^3 \wedge \dd B^1
- a_3 B^1 \wedge \dd B^2.   
\eea

In performing these steps, we have dualised the gauge fields to scalars
\be
\label{DDY}
\DD Y_{I} \equiv  \dd Y_I +C_{IJK} a_J B^K  = e^{2 W_{I}} *_3 G^I, 
\ee
and introduced complex coordinates, $z_I = e^{W_I} + i Y_{I}$, where $g_{I \bar{J}} = \partial_{I} \partial_{\bar{J}} K$ corresponds to the K\"{a}hler metric. $C_{IJK}$ denote constants that are symmetric in the indices, i. e. $C_{IJK} = |\epsilon_{IJK}|$. The potential has been elegantly recast in terms of $T$ and its derivatives. We note that the scalar manifold is [SU(1,1)/U(1)]$^3$. 

With the introduction of $T$, the task of identifying supersymmetric AdS$_3$ vacua is immediate; vacua correspond to critical points of $T$, $\partial_{W_I} T =0$, \cite{Karndumri:2013dca} 
\bea
\label{ads_crit}
e^{W_I} = - \frac{\prod_{J\neq I} a_J}{\kappa +2 a_I}, 
\eea
and for generic $a_I$, the AdS$_3$ vacua are dual to two-dimensional $\mathcal{N} = (0,2)$ SCFTs. 
As we shall demonstrate later, extremising $T$ is equivalent to solving the Killing spinor equations to find AdS$_3$ vacua, an approach adopted in \cite{Benini:2013cda, Maldacena:2000mw}. Given a knowledge of $T$, it is easy to extract AdS$_3$ vacua. For example, one quickly recognises that there is no AdS$_3$ vacuum when two of the constants $a_I$ vanish and supersymmetry is enhanced to $\mathcal{N} = (4,4)$. This is a curious feature, since the near-horizon of D1-D5-branes supports such an AdS$_3$ vacuum and its absence may be attributed to the non-compactness of the target space \cite{Maldacena:2000mw}.  When one of the $a_I$ are set to zero and supersymmetry is enhanced to $\mathcal{N} = (2,2)$ - for concreteness $a_3 =0$ -  solving $\partial_{W_I} T = 0$, we find the equations
\bea
e^{W_3} &=& \frac{a_1}{2} = \frac{a_2}{2}, \quad a_1 e^{-W_2} + a_2 e^{-W_1} = 2.  
\eea
Combined with (\ref{susy_cond}), one quickly sees that $\kappa < 0$, i. e. that the Riemann surface is necessarily hyperbolic. As a consequence of this observation, we remark that the theories studied by Almuhairi-Polchinski \cite{oai:arXiv.org:1108.1213} require $a_I \neq 0$. Moreover, we note that $W_1$ and $W_2$ have only a single constraint, so there is a class of marginal deformations of the theory \cite{Maldacena:2000mw}.  

\section{Supersymmetry conditions}
\setcounter{equation}{0}
\label{sec:susy}

To find all the supersymmetric solutions of 3D U(1)$^3$ gauged supergravity, we require a knowledge of the Killing spinor equations. To deduce these, we can either perform a dimensional reduction of higher-dimensional fermionic supersymmetry variations, a  procedure that serves to pin-down the exact identity of a lower-dimensional bosonic theory. Alternatively, given the bosonic sector of the reduced theory, it is possible to reconstruct the fermionic sector and extract the Killing spinor equations. This latter approach was adopted in \cite{Colgain:2010rg} for 3D $\mathcal{N} =2$ gauged supergravity \cite{deWit:2003ja}. For completeness, here we will perform both.  

We recall that the embedding of our theory in 5D U(1)$^3$ gauged supergravity is understood, so we begin in 5D and reduce the fermionic supersymmetry variations, which, once set to zero, will present us with our desired Killing spinor equations. This task was partially completed in \cite{Karndumri:2013dca}, where it was noted that for supersymmetric AdS$_3$ vacua, the process of solving the Killing spinor equations in 5D and extremising the 3D superpotential should be equivalent. We will complete the task here and confirm that this is indeed the case. 

We follow the conventions of \cite{Behrndt:1998ns} (see also \cite{Maldacena:2000mw}). We recall that the 5D U(1)$^3$  theory \cite{Cvetic:1999xp} consists of three gauge fields $A^{I}$, with field strengths $F^{I} = \dd A^{I}$, and three constrained scalars $X^{I}$, $I=1, 2, 3$, $X^1 X^2 X^3 =1$, which may be further expressed in terms of two scalars $\varphi_i$, $i =1, 2$:
\be
\label{constrained_scalars}
X^1 = e^{-\frac{1}{2} (\frac{2}{\sqrt{6}} \varphi_1 + \sqrt{2} \varphi_2 )}, ~~X^2 = e^{-\frac{1}{2} (\frac{2}{\sqrt{6}} \varphi_1 - \sqrt{2} \varphi_2 )}, ~~X^3 = e^{\frac{2}{\sqrt{6}} \varphi_1}
\ee

In 5D, the fermionic supersymmetry variations may be written as \cite{Behrndt:1998ns}
\bea
\label{5d_diff}
\delta \psi_M &=& \left[ \nabla_M + \frac{i}{24} (X^I)^{-1} ( \Gamma_{M}^{~NP} - 4 \delta_{M}^{~N} \Gamma^P ) F^I_{NP} + \frac{1}{2} V_{I}X^I \Gamma_M- \frac{i}{2}  \sum_{I} A^I_{M} \right]  \epsilon, \\
\label{5d_alg}
\delta \chi_{(i)} &=&  \left[ \frac{1}{8} \partial_{\varphi_i} (X^I)^{-1} F^I_{MN} \Gamma^{MN}  + \frac{i}{2} \partial_{\varphi_i}  \sum_{I} X^I - \frac{i}{4} \delta_{ij} \partial_{M} \varphi_j \Gamma^M  \right] \epsilon, 
\eea
where $\nabla_{M} \equiv \partial_{M} + \frac{1}{4} \omega_{M}^{~NP} \G_{NP}$, $V_{I} = \frac{1}{3}$ and it is understood that repeated indices are summed. 

We will now perform a dimensional reduction on a genus $\frak{g}$ Riemann surface $\Sigma_{g}$ by considering an ansatz of the form: 
\bea
\label{red_ansatz}
e^{a} &=& e^{-2C} \bar{e}^{a}, \quad e^{\alpha} = e^{C} \bar{e}^{\alpha}, \nn
F^{I} &=& G^{I} - a_{I} \vol (\Sigma_{\frak{g}}), 
\eea
where $G^{I}$ is now the field strength for a purely 3D potential, $G^{I} = \dd B^I$ and $a_{I}$ are constants, which correspond to twist parameters in the dual field theory \cite{Bershadsky:1995vm, Vafa:1994tf}. In the choice of ansatz for the frame, $a, b = 0, 1, 2,$ label 3D spacetime directions, $\alpha = 1, 2,$ denote directions
along $\Sigma_{\frak{g}}$ and the scalar warp factor has been chosen to arrive at 3D Einstein frame. The 5D scalars, $\varphi_i$, simply reduce to 3D scalars and the quoted scalars in the 3D gauged supergravity, $W_I$,  are related to these scalars through (\ref{scalars3D5D}). 

In addition to the above ansatz for the bosonic sector of the theory, to perform the reduction we must also specify an ansatz for the supersymmetry parameter, fermions and the 5D gamma matrices,
 \bea 
 \e &=& e^{\beta C} \xi \otimes \eta, ~~ \delta \chi_{(I)} = e^{\beta C} \delta\tilde{\chi}_{(I)} \otimes \eta, \nn
 \delta \psi_{a} &=& \delta \tilde{\psi}_{a} \otimes \eta, ~~~~\delta \psi_{\alpha} = e^{\beta C} \delta \tilde{\chi}_{(3)} \otimes \sigma_{\alpha} \eta, \nn
 \Gamma^{a} &=& \gamma^{a}
\otimes \s^3, ~~~\Gamma^{\alpha+2} = 1 \otimes \s^{\alpha}, 
 \eea
 where in the last line we have made use of the Pauli matrices to decompose the gamma matrices. 
In the reduced theory, $\xi$ corresponds to the 3D supersymmetry parameter,  namely the Killing spinor, while (dropping tildes) $\chi_{(I)}, I = 1, 2, 3$ denote linear combinations of three spinor fields and a (complex) gravitino $\delta \psi_{a}, a= 0, 1, 2$, as we will see in due course. $\eta$ denotes a constant spinor on the Riemann surface satisfying  $\sigma^3 \eta = \eta$. We have introduced the constant $\beta$ for later convenience.  
 
Decomposing the 5D algebraic fermionic variations (\ref{5d_alg}), we get 
\bea 
\label{phi1} \sqrt{6} \delta \chi_{(1)} &=& \biggl[
\frac{1}{8}\sum_{I=1}^2 (X^I)^{-1} \left( e^{4C} G^I_{a b} \g^{a
b} - 2i a_I e^{-2C} \right) - \frac{1}{4} (X^3)^{-1} \left( e^{4C}
G^3_{ab}
\g^{ab}  - 2i a_3 e^{-2C} \right) \nn& & +\frac{i}{2} \left(-X^1 - X^2 + 2 X^3\right) - i \frac{\sqrt{6}}{4} e^{2C} \partial_{a} \varphi_1 \gamma^{a} \biggr]  \xi, \\
\label{phi2} \sqrt{2} \delta \chi_{(2)} &=& \biggl[
\frac{1}{8} (X^1)^{-1} \left(e^{4C} G^1_{ab} \g^{a
b} - 2i a_1 e^{-2C} \right) -\frac{1}{8} (X^2)^{-1} \left(e^{4C} G^2_{ab} \g^{a
b} - 2i a_2 e^{-2C} \right) \nn && +\frac{i}{2} \left(-X^1 + X^2
\right) - i \frac{\sqrt{2}}{4} e^{2C}
\partial_{a} \varphi_2 \gamma^{a} \biggr]  \xi. 
\eea
We find an additional algebraic contribution to the 3D spinor field variations from the differential fermionic variation (\ref{5d_diff}) along $\Sigma_{\frak{g}}$,  
\bea
\label{phi0} 2 
\delta \chi_{(3)} &=& \biggl[\g^{a} e^{2C} \partial_{a} C +
\sum_{I=1}^3 \biggl( \frac{1}{3} X^I  - \frac{1}{3} e^{-2C}
a_I (X^I)^{-1} + \frac{i}{12}
e^{4C} \g^{ab} (X^I)^{-1} G^i_{ab} \biggr) \biggr] \xi. 
\eea
To get this expression, one has to impose the supersymmetry condition (\ref{susy_cond}). As a consistency check at this stage, it is possible to see that the expressions vanish when the scalars are set to their AdS$_3$ values (\ref{ads_crit}).

Taking various linear combinations, and making use of the scalar redefinition (\ref{scalars3D5D}), one can rewrite the spinor field variations as (appendix C of \cite{Karndumri:2013dca}) 
\bea
 \delta \tilde{\chi}_{(1)} &=& \left[ \gamma^{a} \partial_{a} W_1 + \frac{i}{2} e^{W_1}  G^1_{ab} \gamma^{ab}  + e^{-W_1} \left( 2 - a_2 e^{-W_3} -a_3 e^{-W_2}\right) \right]
\xi, 
\eea
where we have for the moment suppressed cyclic terms, i. e. $1 \rightarrow 2 \rightarrow 3 \rightarrow 1$. 

Once again making use of (\ref{5d_diff}), we can identify the 3D gravitino variation: 
\bea
\delta \psi_a &=& e^{\b C} \biggl[ e^{2C} \nabla_{a} + \beta e^{2C} \partial_{a} C - e^{2C} \gamma_{a}^{~b} \partial_{b} C + \frac{i}{24} e^{4 C} (X^I)^{-1} \gamma_{a}^{~bc} G^I_{ bc} \nn
&-& \frac{i}{6} e^{4C} (X^I)^{-1} \gamma^{b} G^I_{a b} + \frac{1}{12} e^{-2C} a_I (X^I)^{-1} \gamma_{a}  + \sum_I \left( \frac{1}{6} \ X^I \gamma_{a} - \frac{i}{2} e^{2C} B^I_{a} \right) \biggr]  \xi, 
\eea
where repeated $I$ indices are summed. Contracting this expression with $\gamma^{c}$, taking $\beta = -1$ and absorbing warp factors, we can rewrite this as 
\bea
\delta \psi_{a} &=& \biggl[ \nabla_{a} + \frac{1}{2} e^{-2C} \sum_{I} X^I  \gamma_{a} - \frac{1}{4} a_I (X^I)^{-1} e^{-4C} \gamma_{a} + \frac{i}{8} (X^I)^{-1} e^{2C} \gamma_{a}^{~bc} G^I_{bc} - \frac{i}{2} \sum_{I} B^I_{a} \biggr] \xi.  \nonumber
\eea
This completes our reduction of the fermionic supersymmetry variations in an admittedly unshapely form. To make sense of the variations and elucidate the underlying supersymmetric structure, it is advantageous to make use of the superpotential (\ref{T}). Using $T$, the supersymmetry variations may be elegantly recast as 

\bea
\label{KSeq1}
\delta \psi_{a} &=& [ \mathcal{D}_{a} + T \gamma_{a} + \frac{i}{8} \sum_{I} e^{W_I} \gamma_{a}^{~bc} G^I_{bc}  ] \xi, \\
\label{alg0}
\delta \chi_{(I)} &=&  [\gamma^{a} \partial_{a} W_I + \frac{i}{2} e^{W_I} G^I_{ab} \gamma^{ab} - 4 \partial_{W_I} T ] \xi, 
\eea
where we have defined the derivative $\mathcal{D}_{a} \equiv \nabla_a - \frac{i}{2} \sum_{I} B^I$ and in contrast to previous expressions, repeated indices are not summed.  One can check that $\delta \chi_{(I)} =0$ when $\partial_{W_I} T = B^I = 0$ and that one recovers the expected Killing spinor equation for AdS$_3$ with radius 
\be
\label{ell}
\ell =  \frac{2 a_1 a_2 a_3}{2 (a_1 a_2 + a_3 a_1 + a_2 a_3) - a_1^2 - a_2^2 -a_3^2}. 
\ee
Through the usual holographic prescription \cite{Brown:1986nw}, $c= \frac{3 \ell}{2 G_3}$,  one can derive the correct central charge $c$. Since $\ell = \frac{1}{2 T}$ at the AdS$_3$ critical point, one can also extract $c$ from extremising $T^{-1}$ \cite{Karndumri:2013iqa}.  

Now that we have derived the supersymmetry variations of the 3D supergravity, we check that they fall into the expected form of a gauged supergravity. It has already been noted \cite{Karndumri:2013iqa}, that this is the case for the bosonic sector.  A similar exercise was performed in \cite{Colgain:2010rg} and the similarities are quite strong with the K\"{a}hler scalar manifold involving (products of) the hyperbolic space, H$^2$, once we can ignore the contribution from a holomorphic superpotential. Such a term is precluded once the SO(2) R symmetry is gauged, which is the case at hand. 

From \cite{deWit:2003ja}, we know for $\mathcal{N} =2$ supersymmetry that the superpotential $T$ can be expressed quadratically in terms of moment maps $\mathcal{V}^{I}$ and a symmetric embedding tensor $\Theta_{IJ}$ encoding the gauged isometries: 
\be
T = 2 \mathcal{V}^{I} \Theta_{IJ} \mathcal{V}^J.  
\ee
Once isometries are gauged, the partial derivatives in the kinetic terms for the scalar manifold are upgraded to covariant derivatives and the action picks up Chern-Simons terms that are also fixed by the embedding tensor 
\be
\mathcal{L}_{\textrm{CS}} = \frac{1}{2} \mathcal{A}^{I} \Theta_{IJ} \dd \mathcal{A}^J. 
\ee
Note that here we have restricted ourselves to Abelian gaugings. To make comparison, we now set $\mathcal{A}^I = B^I, I = 1, 2, 3$ and adopt the following 
\bea
\mathcal{V}^0 &=& 1, \quad \mathcal{V}^I = \frac{1}{4} e^{-W_I}, \nn
\Theta_{I0} &=& \frac{1}{2}, \quad \Theta_{IJ} = -C_{IJK} a_K. 
\eea
Here $\mathcal{V}^0=1$ corresponds to a central extension of the isometry group and generates the SO(2) R symmetry. It is easy to check that this choice recovers the Chern-Simons term and the superpotential $T$. Adopting a complex gravitino, $\psi_{\mu} = \psi^1_{\mu} + i \psi^2_{\mu}$, and complex spinor $\xi = \epsilon_1 + i \epsilon_2$, we can write the fermionic supersymmetry variations as \cite{deWit:2003ja} (see also \cite{Colgain:2010rg})  \bea
\label{KSE1} \delta \psi_{\mu} &=& \mathcal{D}_{\mu} \xi + \frac{i}{4} \sum_{I} e^{-W_I} \DD_{\mu} Y_I \xi +T \gamma_{\mu} \xi \nn
\label{KSE2} \delta \lambda^I &=& \gamma^{\mu} \DD_{\mu} z^I \xi - 4 e^{W_I} \partial_{W_I} T \xi, 
\eea
where we have defined $\DD z^{I} = \dd e^{W_I} + i \DD Y$, where an expression for $\DD Y^{I}$ can be found in (\ref{DDY}). 

Up to the rescaling $\delta \lambda^{I} = e^{W_I} \delta \chi_{(I)}$, we notice that the supersymmetry variations agree. We also note that  $\DD Y_I = *_3 e^{2 W_I} G^I$, an identity that also follows also from the $Y_{I}$ EOM in the bosonic action.  In the next section, we show that the equations of motion that follow from varying the bosonic action are also a by-product of integrability of the Killing spinor equations, thus confirming that the bosonic and fermionic reductions perfectly match.  

\section{Integrability} 
\setcounter{equation}{0}
\label{sec:integrability}
Given the bosonic action (\ref{Einsteinact}), one can vary the action to derive the equations of motion. The purpose of this section is to show that these EOMs are consistent with the integrability conditions following from the Killing spinor equations. This confirms we have matched the bosonic and fermionic sectors correctly. 

Writing the Killing spinor equation (\ref{KSeq1}) as 
\be
\mathcal{D}_a \xi =  - T \gamma_a \xi - \frac{i}{8} e^{W_I} G^I_{bc} \gamma_{a}^{~bc} \xi, 
\ee
we can act with $\gamma^a \mathcal{D}_a$ on the respective algebraic conditions (\ref{alg0}). For concreteness, we consider 
\be
\delta \lambda^1 = 
 \left[ \gamma^a \partial_a W_1 + \frac{i}{2} e^{W_1} G^1_{ab} \gamma^{ab} - 4 \partial_{W_1} T \right] \xi. 
\ee
We find 
\bea
\label{int_dilatino}
\gamma^a \mathcal{D}_a (\delta \lambda^1) &=& \left[ E_{W_1} + i \gamma^{abc} e^{W_1} (B_{G^1})_{abc}  + i \gamma^{a} e^{-W_1} (E_{B_1})_{a}  \right] \xi
\eea
where we have used 
\bea
(B_{G^I})_{abc} &=& \partial_{[a} G^I_{bc]}, 
\eea
to denote the Bianchis and 
\bea
(E_{g})_{ab} &=& R_{ab} -  \frac{1}{2} \partial_a W^I \partial_b W^I - \frac{1}{2} e^{2 W_I} G^I_{a c} G^{I~c}_{~b} + \frac{1}{4} g_{ab} e^{2 W_I} G^I_{cd} G^{I cd} +g_{ab} [8 T^2 - 8 (\partial_{W_I} T)^2], \nn
E_{W_I} &=& \nabla^2 W_I - \frac{1}{2} e^{2 W_I} G^I_{ab} G^{I \, ab} + \partial_{W_I} [ 8 T^2 - 8 (\partial_{W_K} T)^2  ], \nn
(E_{B^I})_a &=& \nabla_{b} (e^{2 W_I}  G^{I \, b}_{~~a} ) - C^{I}_{~JK} \frac{a_J}{2!} \epsilon^{bc}_{~~a} G^K_{bc}, 
\eea
 the EOMs. To recover the Einstein equation, we make use of the identify 
 \be
\nabla_{[a} \nabla_{b]}\xi = \frac{1}{8} R_{ab c_1 c_2} \g^{c_1 c_2} \xi,  
 \ee
 which when contracted with $\g^b$, and using the the Bianchi $R_{a [c_1 c_2 c_3]}  = 0$ on the RHS, gives 
 \be
 \gamma^b \nabla_{[a} \nabla_{b]}  \xi = - \frac{1}{4} R_{ab } \gamma^b \xi. 
 \ee
 It is easier to rewrite the gravitino variation as
 \be
 \tilde{\mathcal{D}}_{a} \xi = (\nabla_a + A_a + T \gamma_a) \xi, 
 \ee
 where $A_a = \frac{i}{4} e^{-W_I} (e^{2W_I} *_3 G^i )_{a}- \frac{i}{2} (B^1 + B^2 + B^3)_{a}$. 
 We can then deduce that 
 \bea
 \label{int_gravitino}
  \gamma^b \tilde{\mathcal{D}}_{[a} \tilde{\mathcal{D}}_{b]} \xi &=& - \frac{1}{4} (E_{g})_{ab} \gamma^b \xi   - \frac{i}{8} \gamma^b \epsilon_{ab}^{~~c} \sum_{I} (E_{B^I})_c \xi \nn 
  &+& \frac{1}{2} \gamma_{a} \partial_{W_I} T \delta \lambda_I- \frac{1}{8} \partial_a W_I \delta \lambda_I  + \frac{i}{16} e^{W_I} G^I_{bc} \gamma_{a}^{~bc} \delta \lambda_I,  
 \eea
 where repeated indices are summed. This shows that the EOMs derived from the bosonic action are consistent with the Killing spinor equations extracted from the dimensional reduction of the fermionic supersymmetry variations, so that dimensional reduction has been performed correctly for both the bosonic  and fermionic sector, confined to the supersymmetry variations. In principle, one could use the above relations to show that (components of) the Einstein equations are implied once the EOMs for the scalars and gauge fields are satisfied. However, given that we are working in 3D, it is easier to explicitly check the EOMs for the supersymmetric solutions we identify. As we perform this task in the appendix, this makes further analysis here redundant, so we omit it. 

\section{Classification} 
\label{sec:classify}
\setcounter{equation}{0}
We are now in a position to undertake a classification of all supersymmetric solutions. We will use the existence of the 3D Killing spinor as a means to construct spinor bilinears that allow us to convert the Killing spinor equations into differential conditions on the geometry. This will enable us to find all the supersymmetric solutions of 3D U(1)$^3$ gauged supergravity. At this stage, this technique is pretty standard and we refer the unacquainted reader to the original work \cite{Tod:1983pm} and elegant examples in 5D \cite{Gauntlett:2002nw,  Gutowski:2004yv, Gauntlett:2003fk}, which served to popularise the technique. 

Before proceeding, we also remark that our analysis of both timelike and null spacetimes here is implicitly covered by the results of refs. \cite{Gutowski:2004yv} and \cite{Gutowski:2005id}, respectively. From the outset, if our goal was merely to find solutions, we were in a position to introduce a Riemann surface $\Sigma_{\frak{g}}$ directly in 5D. However, the analysis in the earlier sections has helped confirm the correct 3D supergravity structure of the theory and here we opt to follow the classification through in 3D. We outline the connection for the timelike case in section \ref{sec:5Dconn}, thus providing a consistency check on some of the results of ref. \cite{Gutowski:2004yv}. 

We now proceed with the classification. To this end, we introduce a set of Killing spinor bilinears
\footnote{A concrete choice for the gamma matrices, which we will employ, is $\gamma_0 = -i \sigma_1, \gamma_1 = \sigma_2, \gamma_2= \sigma_3$. With this choice, we then have the inter-twiners $A= \sigma_1$ and $C = \sigma_2$ and $\gamma_{012} = 1$. Further details are in the appendix.}, 
\be
f = \bar{\xi} \xi, \quad i P^{0}_a = \bar{\xi} \gamma_{a} \xi, \quad P^1_a+ i P^2_a = \bar{\xi}^c \gamma_a \xi, 
\ee
comprising one scalar, $f$, one real vector, $P^0$, and one complex vector $P^1+i P^2$.  Acting with the Killing spinor equation (\ref{KSeq1}) on $P^0$, it is easy to show that it satisfies the Killing equation $\nabla_{(a} P^{0}_{b)}=0$, so $P^0$ corresponds to a Killing direction. Making use of the Fierz identity (\ref{Fierz}), one can show that $|P^0|^2 = - f^2$, so $P^0$ is a timelike isometry when $f$ is non-zero, otherwise it is a null Killing vector.  More generally, $P^a \cdot P^b = f^2 \eta^{ab}$, where $\eta^{ab} = (-1, 1, 1)$. 

Before proceeding to the differential Killing spinor equation, we can extract the following information from the algebraic conditions: 
\bea
\label{alg1} P^0_{a} \, \partial^{a} W_I  &=& 0, \\
\label{G} f e^{W_I} G^I &=& - 4 \partial_{W_I} T *_3 P^0 + P^0 \wedge \dd W_I,  
\eea
where there is no summation on $I$ in the second line. 
From the differential condition, we find the following equations,  
\bea
\label{minkdiff0} \dd f &=& 0, \\
\label{minkdiff1} \dd P^0 &=& 4 T *_3 P^0, \\
\label{minkdiff2} e^{- \frac{1}{2} K} \dd [e^{\frac{1}{2} K} (P^1+i P^2) ] &=& (e^{-W_1}+e^{-W_2} + e^{-W_3}) *_3 (P^1+i P^2) \nn &+&i (B^1 +B^2 +B^3) \wedge (P^1 + i P^2). 
\eea
At this point, we immediately see that $f$ is a constant. 
It is also easy to check that the following Lie derivatives vanish \footnote{Here $\mathcal{L}_{K} = i_{K} \dd + \dd i_{K}$ for a Killing vector $K$.  It is easy to check $i_{P^0} \dd W_I = 0$ by simply contracting $P^0$ into (\ref{alg0}), leading to (\ref{alg1}). The same technique works to calculate $i_{P^0} G^I = -f \dd (e^{-W_I})$, which is closed. } 
\be
\mathcal{L}_{P^0} W_I = \mathcal{L}_{P^0} G^I = 0,  
\ee
implying that the vector $P^0$ does indeed generate a symmetry of the solution. The closure of $G^I$ follows from (\ref{alg1}) and (\ref{minkdiff0}). 

\subsection{Timelike case} 
\label{sec:timelike}
We begin by classifying spacetimes with a timelike Killing vector and without loss of generality we normalise $f=1$. Since $P^0$ is Killing, we can locally introduce a coordinate $\tau$, such that $P^{0} = \partial_{\tau}$. As a result, the 3D spacetime metric may be expressed as 
\be
\dd s^2_3 = -(\dd \tau + \rho)^2 + \dd s^2(M_2), 
\ee
where $\rho$ is a one-form connection on a base Riemann surface, $M_2$, satisfying $\dd \rho = 4 T \vol(M_2)$. From the 5D perspective, this introduces a second Riemann surface in addition to $\Sigma_{\frak{g}}$, which allows us to uplift our results to 5D. Since $P^0$ has been shown to be a symmetry of the entire solution, $\rho$ only depends on $M_2$. From (\ref{G}), it is then easy to convince oneself that the gauge potential for $G^I$, $B^I$ takes the form 
\be
\label{B}
B^I = e^{-W_I} P^0 + \tilde{B}^I, 
\ee
where $\tilde{B}^I$ only depends on $M_2$. At this point, we can use the equation of motion for $G^I$, namely 
\be
\dd (e^{2W_1} * G^I) =C_{IJK} a_J G^K . 
\ee
Taking the Hodge dual of (\ref{G}), multiplying by $e^{W_{I}}$ and differentiating, one finds an equation for the scalar: 
\bea
\label{WI} \dd [ *_2 \, \dd (e^{W_I})]&=& -2\left[4 T - e^{K} \sum_{J\neq I} a_J (e^{W_J}+e^{W_I}) +  e^{K} \prod_{J \neq I} a_J \right] \vol(M_2), 
\eea
where the Hodge dual is now with respect to the metric on $M_2$. Although this equation is second order, in contrast to the equations of motion for $W_{I}$, $G^I$ does not appear and it allows us in principle to determine $W_I$ once we introduce a metric for $M_2$. 

Since $P^1$ and $P^2$ both have unit norm, we can introduce coordinates $x_1, x_2$ through 
\be
P^1 + i P^2 = e^{D-\frac{1}{2} K} (\dd x_1 + i \dd x_2),  
\ee
where $D$, like $K$, is just a function of $x_1$ and $x_2$. We can use the identity 
\be
*_3 (P^1 + i P^2) = -i P^0 \wedge (P^1+i P^2), 
\ee
an expression that can be derived from Fierz identity, to confirm that all $P^0$ dependence drops out of the RHS of (\ref{minkdiff2}).  This allows us to determine the linear combination of the gauge fields in terms of $D$: 
\bea
\label{BD}
\sum_{I} \tilde{B}^I  &=& *_2  \dd D. 
\eea
Taking a derivative, we get 
\bea
\label{Deq0}
\dd *_2 \dd D &=& - 4 \sum_I ( e^{-W_I} \partial_{W_I} T + e^{-W_I} T) \vol(M_2). 
\eea

Equations (\ref{WI}) and (\ref{Deq0}) together now determine the overall solution. It is prudent at this stage to confirm that these equations guarantee a solution to the EOMs. At some level, this is expected, since the integrability conditions (\ref{int_dilatino}) and (\ref{int_gravitino}) show that the bosonic equations of motion are consistent with supersymmetry. To ensure that there are no sign or factor problems in the above analysis, we confirm in appendix \ref{sec:EOMcheck} that the EOMs follow. 

\subsection*{Summary}
Supersymmetric timelike spacetimes correspond to a timelike Killing direction fibered over a Riemann surface parametrised by $(x_1, x_2)$. The 3D solution may be expressed as 
\bea
\label{timelike_soln}
\dd s_3^2 &=& -( \dd \tau + \rho)^2 + e^{2D - K} (\dd x_1^2 + \dd x_2^2), \nn
G^{I} &=& e^{-W_I} \left[ -4 \partial_{W_I} T e^{2D-K} \dd x_1 \wedge \dd x_2 + (\dd \tau + \rho) \wedge \dd W_I \right], 
\eea

where $ \dd \rho = 4 T e^{2D-K} \dd x_1 \wedge \dd x_2$, and the scalars, $W_I$ and warp factor of the Riemann surface are subject to the equations: 
\bea
e^{-W_I} \label{WIeq} \nabla^2 e^{W_I} &=& 16 \left[ \sum_{J \neq I} \partial_{W_J} T \partial^2_{W_I W_J} T - T \, \partial_{W_I} T \right] e^{2D}, \\
\label{Deq} \nabla^2 D &=& 4 \sum_I ( e^{-W_I} \partial_{W_I} T + e^{-W_I} T) e^{2D-K}, 
\eea
where we have rewritten the $W_I$ equation to highlight the fact that the supersymmetry conditions only depend on $T$. 
This provides an explicit derivation of the solution and equations first presented in \cite{Colgain:2015ela}. 

\subsection*{Fixed-points} 
From (\ref{WIeq}), we see that in addition to the supersymmetric AdS$_3$ vacuum (\ref{ads_crit}), a second fixed-point (constant $W_I$) exists: 
\be
e^{W_I} = \sum_{J\neq I} a_J + \frac{\kappa}{2} + \frac{\prod_{J \neq I} a_J}{\kappa}. 
\ee
This fixed-point only exists when $\kappa < 0$, which we set to $\kappa=-1$,  and it is real in a particular range of parameter space, details of which can be found in \cite{Colgain:2015ela}. At fixed-points, (\ref{Deq}) reduces to the Liouville equation $\nabla^2 D = - \mathcal{K} e^{2D}$, where $\mathcal{K}$ is the Gaussian curvature of the Riemann surface $M_2$. At AdS$_3$ fixed-points, where $G^{I} =0$, one can solve the Liouville equation to recover global AdS$_3$ with radius $\ell$ (\ref{ell}), as expected.  

Introducing a radial direction $r = \sqrt{x_1^2 + x_2^2}$ for the Riemann surface and a U(1) isometry $\varphi$, given the Guassian curvature at the fixed-point, $\mathcal{K} = 2 (a_1 a_2 + a_2 a_3 + a_3 a_1) - a_1^2-a_2^2-a_3^2$, solutions to the Liouville equation can be written as 
\be
e^{D} = \frac{2 \sqrt{|\mathcal{K}|}}{|\mathcal{K}|+ \mathcal{K} r^2}. 
\ee

Inserting this, along with $\rho$ into the metric, at the new fixed-point, the spacetime reads: 
\bea
\dd s^2_3 &=& - \ell^2 \left(\dd \tau - \frac{ \textrm{sgn}(\mathcal{K}) r^2}{[1 + \textrm{sgn}(\mathcal{K}) r^2 ]} \dd \varphi \right)^2 + \frac{e^{-K}}{|\mathcal{K}|} \left[ \frac{4 ( \dd r^2 + r^2 \dd \varphi^2)}{(1 + \textrm{sgn}(\mathcal{K}) r^2)^2}\right],  
\eea
where we have isolated a (unit radius) constant curvature Riemann surface  in the upper line with curvature $\textrm{sgn}(\mathcal{K})$. The corresponding expression for $G^{I}$ may be worked out from (\ref{timelike_soln}). 

We observe that regions in parameter space where $\mathcal{K} < 0$ correspond to causally G\"{o}del spacetimes, while those with $\mathcal{K} > 0$ can be analytically continued to either a Berger sphere (squashed S$^3$) or warped AdS$_3$ in Euclidean signature. One can get spacelike warped AdS$_3$ by  either reversing the sign of $T$ or analytically continuing it, $T \rightarrow i T$, however this involves either complex fluxes or giving up the embedding in string theory. If one demands our 3D solutions correspond to real vacua of string theory, these possibilities are precluded. 

Points in parameter space with $\mathcal{K} =0$ fixed-points, where supersymmetry is enhanced, are ruled out. As further details can be found in \cite{Colgain:2015ela}, we omit further discussion on the parameter space here, but reproduce Figure 1. of \cite{Colgain:2015ela} to make this work self-contained. We remark that the constants $a_I$ should be quantised so that the geometry is well-defined, leading to the constraint $2 a_I (\frak{g} -1) \in \mathbb{Z}$. For $\frak{g} \leq 1$, this precludes points in the interior region of Figure 1., however as $\frak{g} > 1$, this proves less of an obstacle and we are free to increase the genus to suitably populate the internal region. 

\begin{figure}[h]
\label{fig:vacua0}
\centering
\includegraphics[width=0.5\textwidth]{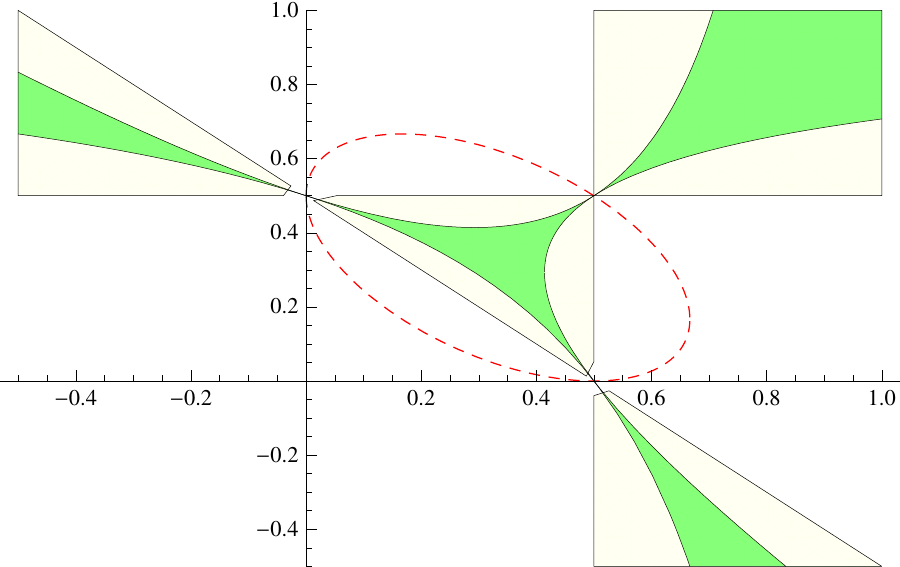}
\caption{The range of good AdS$_3$ vacua (cream) contrasted against points in parameter space (green), where a second fixed-point exists when the genus $\frak{g}$ of $\Sigma_{\frak{g}}$ is greater than one, i. e. when $\Sigma_{\frak{g}} =$ H$^2$. }
\end{figure}

We note that the above metrics all suffer from closed timelike curves (CTCs), since the $g_{\varphi \varphi}$ component of the metric changes sign. One has the freedom to change the connection $\rho$, however CTCs cannot be avoided. Examples are known where oxidation to higher dimensions allows one to exorcise the CTCs \cite{Herdeiro:2000ap} by decompactifying the U(1) direction and going to the covering space of the manifold. This will not work here; our U(1) corresponds to a polar coordinate, so one cannot decompactify it.  Moreover, making use of the uplift of ref. \cite{Cvetic:1999xp}, the requirement that there be no CTCs may be recast as the condition: 
\bea
\frac{4 e^{-K}}{|\mathcal{K}|} \left( r^{-2}+ \frac{16 e^{-K} (\partial_{W_I} T)^2}{|\mathcal{K}|}  \right) \geq \ell^2. 
\eea
We recognise that only at the supersymmetric AdS$_3$ vacuum, where $\partial_{W_I} T =0$, is this condition satisfied, since $r < 1$ for the Poincar\'e disk. For all other spacetimes, the metric flips signature at a given value of $r$. 

\subsection{Null case} 
\label{sec:null}
In this section, we will address the general form of null spacetimes, which are characterised by the Killing vector having zero norm. 
Here $P^0 \wedge *_3 P^0 = 0$ then implies $P^0 \wedge \dd P^0 =0$ through (\ref{minkdiff1}), allowing us to introduce a coordinate $x^+$, such that $P^0 = H^{-1} \dd x^+$ for a given function $H$.  A second implication of the same equation is $P^0 \cdot \nabla P^0 = 0$, so $P^0$ is tangent to affinely parametrised geodesics in the surfaces of constant $x^+$. We can then choose coordinates $(x^+, x^-, r)$, such that 
\be
P^0 = \frac{\partial ~}{\partial x^-}, 
\ee
and the metric takes the form 
\bea
\label{flat_metric} \dd s^2_3 &=& 2 e^+ e^- + (e^3)^2, \\
&=& H^{-1} \left[ \mathcal{F} (\dd x^+)^2 + 2 \dd x^+ \dd x^- \right] + H^2 \dd r^2, 
\eea
where we have introduced a natural orthonormal frame: $e^+ = H^{-1} \dd x^+,\, e^{-} = \dd x^{-} + \frac{1}{2} \mathcal{F} \dd x^+, \, e^{r} = H \dd r$, where $H$ and $\mathcal{F}$ are only independent of $x^-$.  More generally, the metric may also have $g_{+ r}$ terms, but one can make use of a coordinate transformation $r \rightarrow r'(x^+, r)$ to eliminate these, so we have dropped them. The same transformation also serves to rescale the $g_{rr}$ component of the metric. 

At this point, given we have a single underlying spinor $\xi$ with two complex components, it makes sense to also work explicitly with it: 
\be
\xi = \left( \begin{array}{c} \alpha_1 \\ \alpha_2 \end{array} \right), 
\ee
where $\alpha_i \in \mathbb{C}$. We further redefine the gamma matrices
\be
\gamma_{\pm} = - \frac{1}{\sqrt{2}} ( \gamma_1 \pm \gamma_0), 
\ee
such that $\{\gamma_a, \gamma_b \} = 2 \eta_{ab}$, where $\eta_{ab}$ is the metric given in (\ref{flat_metric}). In addition to $f = \bar{\xi} \xi = 0$, aligning $P^0$ with $e^+$ constrains the spinor so that $\alpha_1 = 0$. As a direct consequence, we see that 
\be
\label{projection}
\gamma^+ \xi = 0, 
\ee
so all our solutions preserve half the supersymmetry. Without loss of generality, we will now take $P^0 = e^{+} = H^{-1} \dd x^{+}$. To do this consistently, one has to redefine $H$ to absorb the norm of $\alpha_2$, thus leaving two real components. 

From the algebraic Killing spinor equation (\ref{alg0}), it is straightforward to see that $i_{P^0} G^{I} =0$. As a result, $G^I$ have only components $G^{I}_{+ r}$ and through a gauge transformation $B^{I} \rightarrow B^{I} + \dd \Lambda^{I}(x^{+}, r)$, we can further simplify by setting $B^{I}_{r} =0$. One finds that (\ref{minkdiff1}) is satisfied provided
\be
\label{null_susy1}
\partial_{r} H^{-1} = 4 T. 
\ee 
This condition also imposes the vanishing of $\delta \psi_{r}$ (\ref{KSeq1}), once (\ref{projection}) is imposed, and provided $\partial_{r} \xi =0$. From the vanishing of $\delta \chi_{(I)}$ (\ref{alg0}), or alternatively from (\ref{G}), we find
\be
\label{null_susy2}
- \partial_{r} W_I = 4 \partial_{W_I} T H. 
\ee
We observe that this equation tells us that null spacetimes with constant $W_I$ only exist at the supersymmetric AdS$_3$ critical point. 

By combining these two equations, we can show that (\ref{minkdiff2}) is satisfied. To appreciate this fact, we determine the bilinear 
\be
(P_1 + i P_2)_{+} =  \bar{\xi}^{c} \gamma_{+} \xi = e^{2 i \beta}, 
\ee
where $e^{i \beta}$ is simply the phase of $\alpha_2$ spinor component. As we have just seen, this phase is independent of $r$ and drops out (\ref{minkdiff2}), along with $B^{I}$. This equation, then reduces to 
\be
- \frac{1}{2} \sum_{I=1}^3 \partial_{r} W_I - \partial_{r} \log H = \sum_{I=1}^3 e^{-W_I} H, 
\ee
which can be shown to hold using the explicit expression for the superpotential (\ref{T}). We remark that the $\delta \psi_{-}$ variation trivially vanishes once $\partial_{-} \xi =0$. We confirm in appendix \ref{sec:EOMcheck} that the scalar EOM and the Einstein equation along $E_{+-}$ and $E_{rr}$ are satisfied once (\ref{null_susy1}) and (\ref{null_susy2}) hold. 

The final supersymmetry condition to be imposed is $\delta \psi_{+} =0$. This may be rewritten in the form: 
\be
\label{psi_+}
 \partial_{+} \xi =  \frac{i}{4} H^{-1} \sum_{I=1}^3  \left( 2 B^I_{+} +  e^{W_I} G^{I}_{r+}  \right) \xi. 
\ee

We observe that since $\partial_r \xi = 0$, the RHS has to be independent of the radial direction $r$. To see if this is the case, we can introduce functions $g_{I} (r, x^+)$, so that $B^{I} = g_{I} \dd x^+$. The EOMs for the gauge fields can then be written as 
\be
\label{flux_gI}
\partial_{r} \left( H^{-1} e^{2 W_{I}} \partial_r g_I \right) = -C_{IJK} a_J \partial_r g_K. 
\ee
where there is no sum over $I$. Now using the above EOM, (\ref{null_susy2}) and an explicit expression for $T$, it is possible to show that the RHS of (\ref{psi_+}) is independent of $r$, so that the final supersymmetry condition can be consistently solved. We note that when the RHS of (\ref{psi_+}) vanishes, the Killing spinor is independent of $\xi$ and the number of preserved supersymmetries, neglecting enhancement due to twist parameters vanishing, is two. When the RHS does not vanish, $\alpha_2$ is further determined up to a phase, a constraint that results in a single supersymmetry. 

We are this left with the task of imposing the flux EOMs for the gauge fields and the $E_{++}$ component of the Einstein equation. We will then be in a position to determine the $x^{+}$ dependence of the Killing spinor, since it is not fixed by (\ref{minkdiff2}). We can solve the flux EOMs, by introducing functions $g_{I} (r, x^+)$, so that $B^{I} = g_{I} \dd x^+$.  
The remaining Einstein equation then reads
\bea
\label{E++}
- \frac{1}{2} \partial_{+}^2 H - \frac{1}{2} H^{-1} \partial^2_r \mathcal{F} - 4 T \partial_r \mathcal{F} = \frac{1}{2} \sum_{I =1}^3 \left[ H^2 (\partial_{+} W_I)^2 + e^{2 W_I} (\partial_r g_{I})^2 \right].  
\eea

\subsection*{Examples of null spacetimes}
To get a better feel for the null spacetime solutions, it is fitting to consider some examples. The simplest class of null solutions involve interpolating flows from AdS$_5$  on a Riemann surface $\Sigma_{\frak{g}}$ to supersymmetric AdS$_3$ vacua \cite{Benini:2013cda, Maldacena:2000mw, Gutowski:2005id} \footnote{These flows cover static black string solutions, such as those of ref. \cite{Gibbons:1994vm}.}. In this case $\mathcal{F} =0$ and one is left with only (\ref{null_susy1}) and (\ref{null_susy2}) to solve, since all other equations vanish on the assumption that $W_I$ just depend on the radial direction, $r$, and the gauge fields are zero. We note from (\ref{psi_+}) that the Killing spinor, $\xi$, is independent of the coordinates. 

We next consider an example where $W_I$ are constant and at their AdS$_3$ values, as required by (\ref{null_susy2}). We further assume that $\partial_{+} H = 0$, so that one can solve (\ref{null_susy1}) to get
\be
H^{-1} = \frac{2}{\ell} r + c,  
\ee
where $c$ is a constant, which we take to be zero. The spacetime is then 
\be
\dd s^2_3 = \frac{2 r}{\ell} \left[ 2 \dd x^+ \dd x^- + \mathcal{F} (\dd x^+)^2 \right] + \frac{\ell^2}{4 r^2} \dd r^2.  
\ee

We introduce an ansatz 
\be
 g_{I} = \sigma_I r^{\rho}, \quad \mathcal{F} = - \frac{\ell^3 \lambda^2}{2} r^{z-1}
\ee
where $\sigma_I$, $\rho$, $\lambda$ and $z$ are constants. We have chosen $\mathcal{F}$ so as to recover and generalise the results of \cite{Karndumri:2013dca}, where a simpler ansatz was taken. Integrating (\ref{flux_gI}) up to a constant, which we take to be zero since we are considering a radial ansatz, a solution for $\sigma_{I}$ exists provided:  
\be
 2 \prod a_I - \frac{2}{\ell} \sum_{I=1}^3 a_I^2 e^{2 W_I} \rho + 8 \frac{e^{-2 K}}{\ell^3} \rho^3= 0. 
\ee
where it is understood that $W_I$ and $K$ should be evaluated at their AdS$_3$ values. 

From the Einstein equation, we get the following condition: 
\be
\ell^2 \lambda^2 z (z-1) r^{z-2} = \sum_{I=1}^3 e^{2W_I} \rho^2 \sigma_I^2 r^{2 \rho -2}. 
\ee
Once we identify $z = 2 \rho$, this condition becomes algebraic and can be solved for the constant $\lambda$. In turn $\lambda$ can be rescaled to unity by rescaling the coordinates $x^{+}$ and $x^{-}$. 

We now comment on the existence of these vacua when $z=2$, corresponding to $\rho = 1$. When $z=2$, these geometries are equivalent to 3D Schr\"{o}dinger geometries \cite{Son:2008ye, Balasubramanian:2008dm}. The examples we construct here are similar to the 3D Schr\"{o}dinger solutions presented in \cite{Jeong:2014iva}, since both the preserved supersymmetry and the internal geometry is the same. However, in contrast the solutions presented in \cite{Jeong:2014iva}, here the solutions have not been generated via TsT transformations \cite{Lunin:2005jy} and as a consequence, they exist within the consistent truncation ansatz. 

\begin{figure}[h]
\label{fig:vacua1}
\centering
\includegraphics[width=0.6\textwidth]{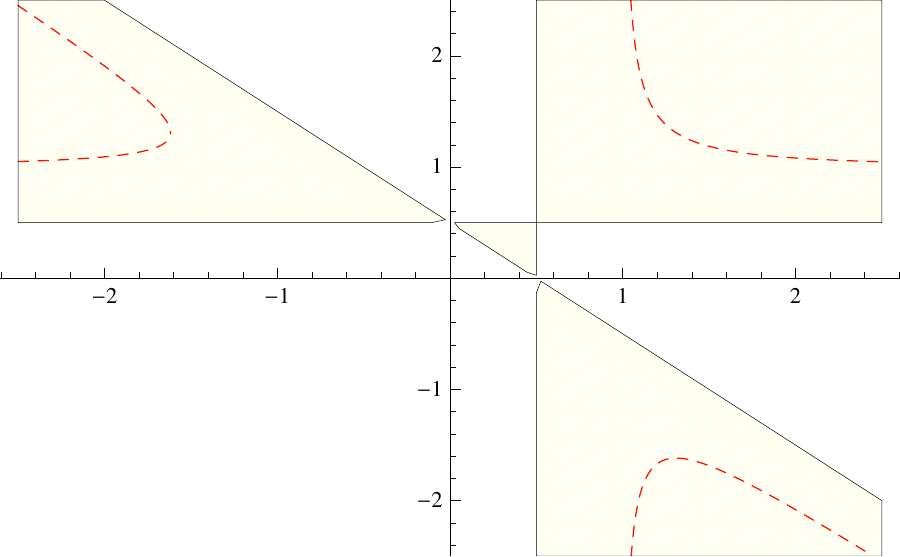}
\caption{The allowed range for good AdS$_3$ vacua contrasted against the loci where null-warped AdS$_3$ vacua exist.}
\end{figure}

We observe that the existence of null-warped AdS$_3$ solutions depends on the Riemann surface $\Sigma_{\frak{g}}$. For example, when $\frak{g}=1$, it is easy to see that one requires $a_I =0$, which is precluded since there is no good AdS$_3$ vacuum for this choice of parameters. Again, for $\frak{g} = 0$, one observes that loci of null-warped AdS$_3$ vacua do not intersect the allowable parameter range (see, for example, Figure 1. of \cite{Benini:2013cda}). On the contrary, when $\frak{g} > 1$ and the internal Riemann surface is a hyperbolic space, we find the null-warped AdS$_3$ vacua can appear when  
\be
a_2 = \frac{-1+ 2 a_3 - a_3^2 \pm \sqrt{a_3 -2 a_3^2 + a_3^4}}{2 (a_3-1)}, 
\ee 
for $a_1 = 1 -a_2 -a_3$, i. e. when the curvature of the hyperbolic space is $\kappa = -1$. From \cite{Colgain:2015ela}, we know this as the special locus in parameter space where no timelike warped AdS$_3$ exist. This is precisely the locus along which G\"{o}del spacetimes become AdS$_3$. 

\section{Supersymmetric flows}
\label{sec:SUSYflow}
\setcounter{equation}{0}
Having introduced the fixed-points in the timelike class, in this section we discuss interpolations by focussing on two illustrative examples. Our intention is not to be exhaustive, but merely to highlight qualitative differences between flows in the interior of Figure 1, namely those where the topology changes, and flows in the external region, where we encounter G\"{o}del fixed-points without topology change.  

We begin with the simplest conceivable example, which corresponds to the most symmetric point in parameter space, i. e. $a_{I} = \frac{1}{3}$ \footnote{Recall that we have normalised the curvature of the internal Riemann surface to unity.}. From the 5D perspective, the fixed-points and interpolating solutions then correspond to solutions to minimal 5D gauged supergravity, which may be uplifted further on a host of supersymmetric geometries to higher dimensions \cite{Gauntlett:2006ai, Gauntlett:2007ma}. In this case, the flow equations required to be solved simplify accordingly,
\bea
\label{W3a_eq} &&\nabla^2 e^{W} = \frac{2}{9} [ 54 e^{2W} - 21 e^{W} + 1] e^{2D}, \\
\label{D3a_eq} &&\nabla^2 D = [ 12 e^{W} -1] e^{2D}, 
\eea
where we have identified the scalars $W_I = W$. We note that the AdS$_3$ critical point, with topology $\mathbb{R} \times$H$^2$ corresponds to $e^{W} = \frac{1}{3}$, while its counterpart with topology $\mathbb{R} \times$S$^2$ appears at $e^{W} = \frac{1}{18}$. By linearising the equations, we immediately recognise that the AdS$_3$ critical point is perturbatively unstable, and it is the second fixed-point that exhibits attractive behaviour.

This instability of AdS$_3$ means that once we choose the initial value of $W_0$ below its AdS$_3$ value, the scalar flows towards the second fixed-point. In this early regime $D$ increases until it hits $W = -\log (12)$, at which point it starts to decrease. In the meantime, $W$ continues on its trajectory, passes through the second fixed-point, before rebounding and starting to oscillate. The oscillations freeze out and the dynamics end when $D$ gets small. It is conceivable that the right initial conditions can be found so that the trajectory finishes at the second fixed-point. We do not investigate this here, but simply demonstrate that one can connect fixed-points using a shooting method. In this particular example this will ultimately lead to a singular flow as when $D$ gets small, $W$ continues on uninterrupted until $e^{W} \rightarrow 0$ and, as a consequence, the superpotential blows up, $T \rightarrow \pm \infty$. 

\begin{figure}[h]
\label{fig:vacua2}
\centering
\begin{tabular}{cc}
\includegraphics[width=0.45\textwidth]{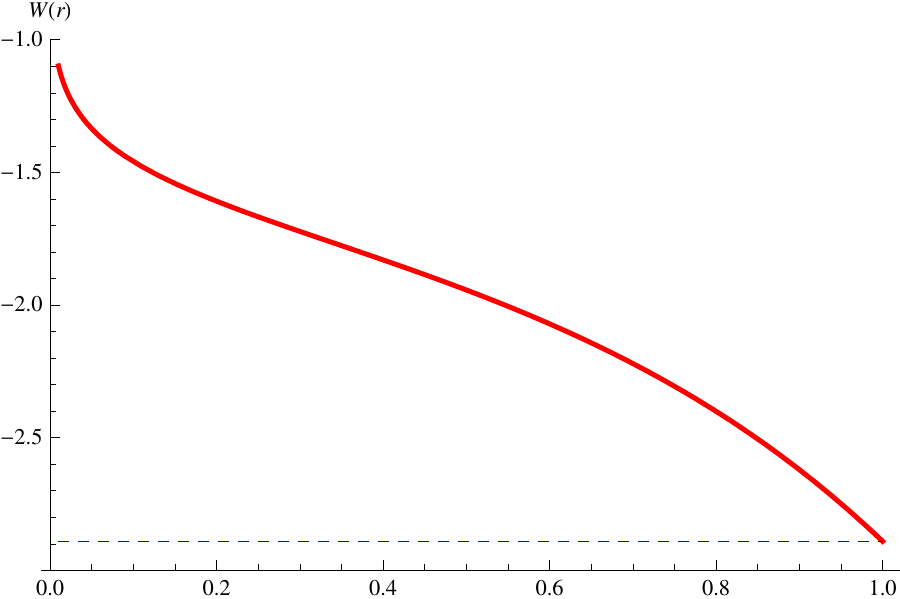} & \includegraphics[width=0.45\textwidth]{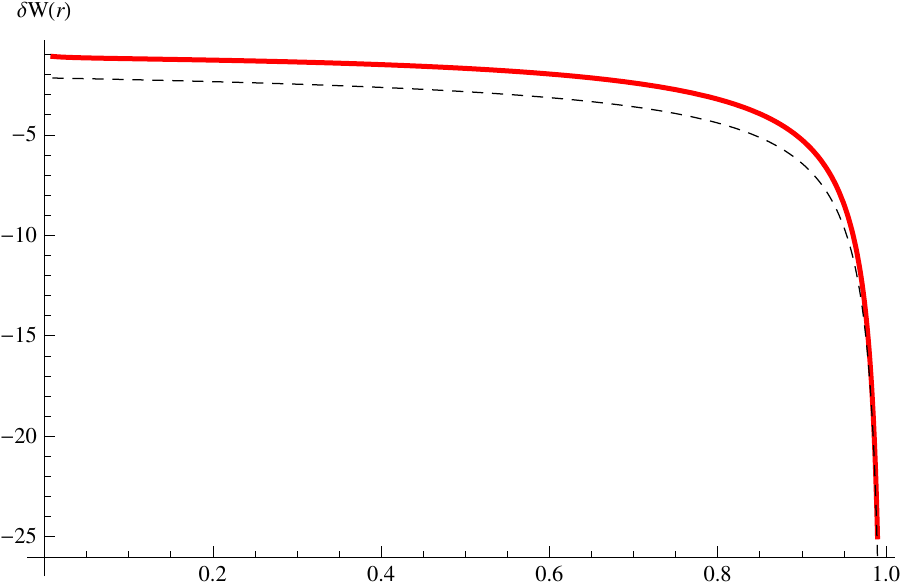} \\
\tiny{(a)} & \tiny{(b)} \\
\end{tabular}
\caption{ (a) shows a (ultimately singular) scalar trajectory connecting the two fixed-points. In (b) we have solved for the corresponding fluctuation in the AdS$_3$ background to show that its behaviour at the boundary corresponds to a non-normalisable mode.}
\end{figure}

Since the supersymmetric AdS$_3$ vacuum is unstable, it is expected that the deformations we have considered to get these flows correspond to deformations of the CFT by an irrelevant operator. We will now determine the conformal dimension of this scalar operator and show that it corresponds to a non-normalisable mode. We start by performing the coordinate transformation 
\be
r =  \frac{1-u^2}{1+u^2}, 
\ee
so that $u$ now corresponds to the customary radial direction of AdS$_3$, with boundary $u=0$. Near the boundary, we therefore have $u \simeq \sqrt{1-r}$. Next we linearise the equation (\ref{D3a_eq}), getting 
\be
\nabla^2 \delta {W} = \frac{1}{r} \delta {W}' + \delta {W}'' = \frac{40}{9} \frac{\delta {W}}{(1-r^2)^2}, 
\ee
where $\delta {W}$ is a fluctuation in the scalar $W$ and derivatives are with respect to $r$. While we also have to consider a fluctuation in the warp factor $D$ to make sure that (\ref{W3a_eq}) is satisfied, it is a pleasing feature that this fluctuation decouples from this equation above. We stress that we are now neglecting the back-reaction of the scalar and simply considering fluctuations in AdS$_3$. Adopting $\delta {W} = (\sqrt{1-r})^p \simeq u^{p}$, we can determine $p$ in the limit $r \rightarrow 1$, where we encounter the AdS$_3$ boundary. Doing so, we find $p = \frac{10}{3}$ and $p  = -\frac{4}{3}$, corresponding to scalar operator with conformal dimension $\Delta = \frac{10}{3}$. In 2D this operator corresponds to an irrelevant operator and by following the flows to the boundary we have confirmed that the non-normalisable mode with $p = - \frac{4}{3}$ is turned on. See Figure 3. (b), where the dashed curve, modulo a suitable coefficient, corresponds to $(1-r)^{-\frac{2}{3}}$. 

The second example we consider is from an external region of Figure 1, where new fixed-points are G\"{o}del spacetimes. For concreteness, we select the point $(a_1, a_2, a_3) = (\frac{3}{2}, \frac{3}{2}, -2)$. This choice will allow us to truncate the theory so that $W_2 = W_1$. With this simplification, the flow equations become: 
\bea
&&\nabla^2 e^{W_1} = 2 [ 2 e^{2 W_1} + 4 e^{W_1+W_3}  - 4 e^{W_1} -3 +4 e^{W_3}] e^{2D}, \nn
&&\nabla^2 e^{W_3} = 2 [ 2 e^{2 W_1} +4  e^{W_1+W_3}  - e^{W_3} +\frac{9}{4} -6 e^{W_1}] e^{2D}, \nn
&&\nabla^2 D = [ 8 e^{ W_1} + 4 e^{W_3} -1] e^{2D}. 
\eea

The AdS$_3$ and G\"{o}del fixed-points are located at $(e^{W_1}, e^{W_3}) = (\frac{3}{2}, \frac{9}{20})$ and $(e^{W_1}, e^{W_3})=(2, \frac{1}{4})$ respectively. In contrast to the previous example, here both fixed-points are perturbatively unstable. By either linearising the above supersymmetry conditions and taking the AdS$_3$ limit, $r \rightarrow 1$, or linearising the scalar EOMs, as we have done in the appendix (\ref{linear_scalar}), one can diagonalise the mass squared matrix to extract the masses, $m^2 \ell^2 = \frac{1}{8}(22 \pm 3 \sqrt{51})$, which correspond to CFT operators of dimensions: 
\be
\Delta = 1\pm \frac{3}{4} + \frac{1}{4} \sqrt{51}. 
\ee
Once again, we see that both correspond to irrelevant operators by following the fluctuation to the AdS$_3$ boundary. Solving the second-order equations numerically, while at the same time choosing the initial conditions in a suitable fashion, it is possible to find flows interpolating between fixed-points, as demonstrated in Figure 4.  We note that this flow is better behaved than the previous example in that $D \rightarrow \infty$ at a given value of $r$. This is a common feature shared with the analytic fixed-point solutions. 
 
\begin{figure}[h]
\label{fig:vacua3}
\centering
\begin{tabular}{cc}
\includegraphics[width=0.45\textwidth]{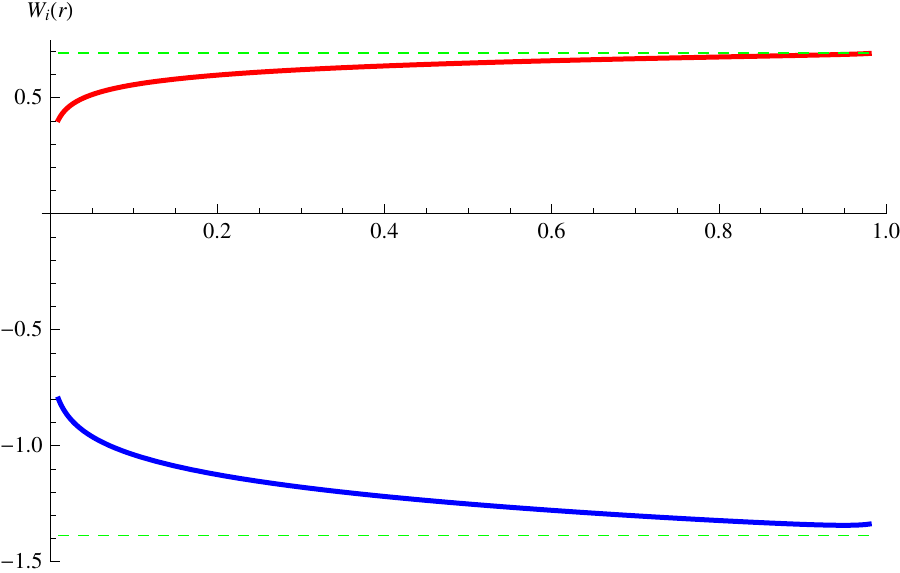} & \includegraphics[width=0.45\textwidth]{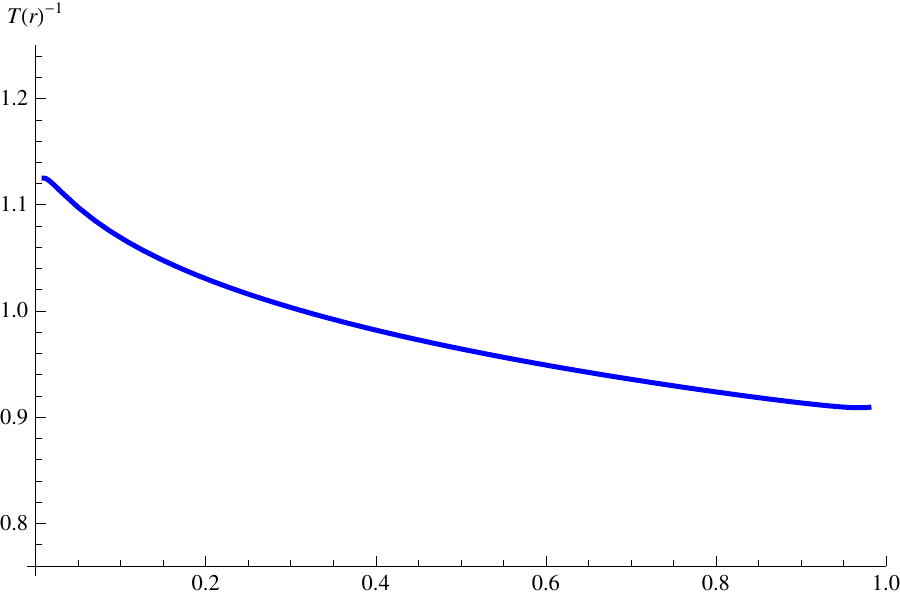} \\
\tiny{(a)} & \tiny{(b)} \\
\end{tabular}
\caption{A flow from AdS$_3$ to the G\"{o}del fixed-point. Dashed green lines denote the values of the scalar at the G\"{o}del fixed-point. We also note that $T^{-1}$ decreases in flowing from AdS$_3$ to G\"{o}del. }
\end{figure}

We remark that $T^{-1}$ appears to play the role of a $c$-function decreasing along the flow, as demonstrated in Figure 4 (b). As pointed out in \cite{Colgain:2015ela}, this suggests that a generalisation of the results of \cite{Compere:2007in} should be considered before applying them to our flows. We hope to explore this direction in future.

\section{Connection to 5D literature} 
\label{sec:5Dconn}
\setcounter{equation}{0}
So far we have been working exclusively in 3D supergravity, and have given little thought to the higher-dimensional realisation of our class of geometries. Here we remedy this and demonstrate that our results for timelike spacetimes are consistent with well-known classifications in 5D \cite{Gutowski:2004yv,Gauntlett:2003fk}. Most relevant is the work of Gutowski-Reall \cite{Gutowski:2004yv}, where timelike solutions of 5D gauged supergravity coupled to arbitrarily many Abelian vector multiplets are presented. Specialising to two vector multiplets, coupled to the graviphoton of the supergravity multiplet, we recover the parent 5D U(1)$^3$ gauged supergravity.   

For completeness, we briefly review  the relevant results of \cite{Gutowski:2004yv}. The 5D metric may be written locally as 
\bea
\dd s^2_5 &=& f^2 (\dd t + \omega)^2 - f^{-1} h_{mn} \dd x^m \dd x^n
\eea
where $f$ is a scalar, $h_{mn}$ denotes the metric on a 4D Riemannian base manifold, $\mathcal{B}$, and  $\omega$ is a one-form connection on $\mathcal{B}$. The two-form $\dd \omega$ splits into self-dual and anti-self-dual parts on $\mathcal{B}$: 
\be
f \dd \omega = G^+ + G^-. 
\ee
The 5D field strength for the gauge fields reads \footnote{To facilitate comparison we have set the coupling to unity, $g = \chi \xi = 1$.}
\bea
\label{flux}
F^{I} &=& \dd [X^{I} f (\dd t + \omega) ] + \Theta^{I} - 9 f^{-1} C^{IJK} V_J X_K  J^{(1)}, 
\eea
where $V_I = \frac{1}{3}$ and $C_{IJK} $ denote constants, with the latter being symmetric in indices. We note that $X^I$ are functions of the unconstrained scalars of the 5D theory (\ref{constrained_scalars}) and satisfy 
\be
\frac{1}{6} C_{IJK} X^{I} X^J X^K =1.  
\ee
One defines $X_I$ so that $X_I \equiv \frac{1}{6} C_{IJK} X^J X^K$.  Completing the expression for $F^I$, we have self-dual two-forms, $\Theta^{I}$, and a closed anti-self-dual two-form, $J^{(1)}$, on $\mathcal{B}$. 

The Ricci-form, $\mathcal{R}_{mn} = \frac{1}{2} J^{pq} R_{pq mn}$, satisfies the following identify 
\be
\label{Ricci}
\mathcal{R} = 3 V_{I} \Theta^{I} - 27 f^{-1} C^{IJK} V_I V_J X_K J^{(1)},  
\ee
and as a direct consequence of the Maxwell equations, we have the equation \footnote{There is a missing $V_{I}$ on the RHS of (2.81) in ref. \cite{Gutowski:2004yv}.}
\bea
\label{Maxwell}
\dd *_4 \dd (f^{-1} X_I ) &=& - \frac{1}{6} C_{IJK} \Theta^{I} \wedge \Theta^J + 2 f^{-1} V_I G^{-} \wedge J^{(1)} \nn 
&+& 6 f^{-2} \left( Q_{IJ} C^{JMN} V_{M} V_{N} + V_{I} X^J V_J \right) \vol(\mathcal{B}), 
\eea
where we have defined $ Q_{IJ} = \frac{9}{2} X_I X_J - \frac{1}{2} C_{IJK} X^K$. These expressions hold for arbitrarily many vector multiplets, but one can specialise to the U(1)$^3$ theory by taking the indices $I, J, K$ to run from 1 to 3 so that $C_{IJK} =1$ if $(IJK)$ is a permutation of $(123)$ and $C_{IJK} = 0$ otherwise.

We will now discuss how our results are related. Firstly, one uplifts the timelike solutions presented in section \ref{sec:timelike} using the consistent truncation identified in \cite{Karndumri:2013iqa}
\bea
\dd s^2_{5} &=& - e^{-4C} ( \dd \tau + \rho)^2 + e^{2C} \left[ e^{2D} ( \dd x_1^2 + \dd x_2^2) + \dd s^2 (\Sigma_{\frak{g}}) \right], \nn
F^{I} &=& - a_I \vol (\Sigma_{\frak{g}}) + G^{I}. 
\eea
Observe that we can analytically continue the 3D coordinates, $\tau, x_1, x_2$, along with connection $\rho$, and the Riemann surface $\Sigma_{\frak{g}}$  to overcome the difference in signature. We further redefine $e^{W_I} = e^{2C} (X^{I})^{-1}$ and one finds the field strength: 
\be
F^{I} = \dd [ X^I f ( \dd \tau + \rho)]  + a_I [ \vol (\Sigma_{\frak{g}}) + e^{2 D} \dd x_1 \wedge \dd x_2 ] - 2 \sum_{J \neq I} e^{W_J} e^{2 D} \dd x_1 \wedge \dd x_2. 
\ee
Relating expressions, $\tau \rightarrow t, \rho \rightarrow \omega$, we get  
\bea
\Theta^{I} &=& \left( a_I - \sum_{J \neq I} e^{W_J} \right) [ e^{2D} \dd x_1 \wedge \dd x_2 + \vol(\Sigma_{\frak{g}}) ] , \nn
J^{(1)} &=& e^{2D} \dd x_1 \wedge \dd x_2 - \vol(\Sigma_{\frak{g}}), \nn
G^{\pm} &=& 2 T e^{4C}  [ e^{2D} \dd x_1 \wedge \dd x_2 \pm \vol(\Sigma_{\frak{g}}) ].  
\eea
Note that this choice of $G^{\pm}$ means that $\omega$ is now a one-form only on the Riemann surface parametrised by $(x_1, x_2)$. As  a final check of consistency, we can recover (\ref{WI}) and (\ref{Deq}) from (\ref{Maxwell}) and (\ref{Ricci}), respectively. Indeed, (\ref{Ricci}) breaks up into two parts and the components along $\Sigma_{\frak{g}}$ neatly recover (\ref{susy_cond}), the condition for supersymmetry. So everything is consistent. 

It would be interesting to see if a more general class of warped dS$_3$ or AdS$_3$ (G\"{o}del) solutions can be found using the results of \cite{Gutowski:2004yv}. Recall that we have reduced the 5D U(1)$^3$ theory on a Riemann surface, so we are confined to direct-product spacetimes, meaning that we only have a one-form connection for one Riemann surface, i. e. $M_2$. Related solutions to 5D minimal gauged supergravity are presented in ref. \cite{Gauntlett:2003fk}, where the base space is a product of Riemann surfaces. The connection on one of the Riemann surfaces degenerates at special points of the parameter $\beta$, in the notation of ref. \cite{Gauntlett:2003fk}, but as can be seen from Figure 1, one is guaranteed to only find the unwarped AdS$_3$ vacuum and a specific example of warped dS$_3$ when $a_I = \frac{1}{3}$. This is the only point of overlap. 

\section*{Acknowledgments}
We thank K. Jensen, P. Karndumri and J. Nian for related discussions. E \'O C is supported by the Marie Curie grant  PIOF-2012-328625 ``T-dualities". 

\appendix 
\section{Conventions}
\setcounter{equation}{0}
We take the conventions for the gamma matrices from \cite{Sohnius:1985qm}. In particular, in three dimensions and signature $\eta_{\mu \nu} = (-1, 1, 1)$, we encounter the following inter-twiners: 
\be
A \gamma_{a} A^{-1} = - \gamma_{a}^{\dagger}, \quad C^{-1} \gamma_{a} C = - \gamma_a^{T}, \quad D^{-1} \gamma_a D = \gamma_a^{*}, 
\ee
where $D = C A^{T}$ and signs are determined by the choice $\gamma_{012} = 1$. Here we are using the fact that since $\gamma_{012}$ commutes with all the other gamma matrices, it is simply proportional to the identity. As it squares to one, the constant of proportionality is $1$. $C$ is anti-symmetric, $C = - C^T$.  

We make use of the following Fierz identify in 3D: 
\be
\label{Fierz}
\bar{\xi}_1 \xi_2 \bar{\xi}_3 \xi_4 = \frac{1}{2} \bar{\xi}_1 \xi_4 \bar{\xi}_3 \xi_2 + \sum_{m=1}^3 \frac{1}{2} \bar{\xi}_1 \gamma_{m} \xi_4 \bar{\xi}_3 \gamma^{m} \xi_2. 
\ee

\section{Equations of Motion} 
\setcounter{equation}{0}
\label{sec:EOMcheck}

\noindent
\textbf{Timelike} \\
In this section, we show explicitly that the Einstein and scalar EOMs for timelike spacetimes are a consequence of our supersymmetry conditions. 

It is an straightforward exercise to check that the scalar equations of motion, namely 
\be
\nabla^2 W_I - \frac{1}{2} e^{2 W_I} G^I_{ab} G^{I\,ab}+ \partial_{W_I} V = 0, 
\ee
are satisfied once (\ref{G}) and (\ref{WI}) hold. 

As for the Einstein equation, 
\bea
\label{Einstein}
R_{ab} &=& \frac{1}{2} \partial_a W^I \partial_b W^I + \frac{1}{2} e^{2 W_I} G^I_{a c} G^{I~c}_{~b} - \frac{1}{4} e^{2 W_I} g_{ab} G^I_{cd} G^{I cd} - g_{ab} V ,  
\eea
where $V = 8 T^2 - 8 (\partial_{W_I} T)^2$, 
a calculation of the Ricci tensor leads to 
\bea
\label{ricci}
R_{00} &=& 8 T^2, \quad R_{11} = R_{22} = 8 T^2 - \nabla^2 (D- \tfrac{1}{2} K) e^{-2 D+ K}. 
\eea
One observes that the Einstein equation in the temporal directions is trivially satisfied once the correct expression for $G^I$ (\ref{G}) is inserted. Given symmetry along the Riemann surface, the remaining Einstein equation can be written as 
\be
[ 16 T^2 -8 (\partial_{W_I} T)^2] e^{2D-K} -\nabla^2 D - \frac{1}{2} \sum_I  e^{-W_I} \nabla^2 e^{W_I} = 0. 
\ee
Using the equations (\ref{WIeq}) and (\ref{Deq}), one can see that this equation is satisfied. \\

\noindent
\textbf{Null} \\
As may be seen by a direct calculation, the scalar EOMs are implied by (\ref{null_susy1}) and (\ref{null_susy2}). We note that if $W_I$ depends on $x^{+}$, it is not fixed by this equation. 

In calculating the Ricci tensor, one can make use of the following spin connections: 
\bea
\omega_{r+} &=& \partial_{+} H e^r - \frac{1}{2}  \partial_r \mathcal{F} e^+ - 2 T e^{-}, \nn
\omega_{r-} &=& -2 T e^{+}, \nn
\omega_{+-} &=& 2 T e^{r}, 
\eea
where again we have made use of (\ref{null_susy1}) and (\ref{null_susy2}). The Ricci tensor is then calculable from $R^{a}_{~b} = \dd \omega^{a}_{~b} + \omega^{a}_{~c} \wedge \omega^{c}_{~b}$, and we find: 
\bea
R_{++} &=& - \frac{1}{2} \partial_{+}^2 H - \frac{1}{2} H^{-1} \partial^2_r \mathcal{F} - 4 T \partial_r \mathcal{F},  \nn
R_{+-} &=& - V =  8 T^2 - 8 \sum_{I=1}^3 (\partial_{W_I} T)^2, \nn
R_{rr} &=& 16 \sum_{I=1}^3 (\partial_{W_I} T)^2  - 8 T^2. 
\eea
It can be shown that the Einstein equations in the $E_{+-}$ and $E_{rr}$ directions are now trivially satisfied. The $E_{++}$ component gives us a final equation (\ref{E++}). 

To identify the mass of the scalar and the corresponding conformal dimensions, it is useful to record the scalar EOM linearised about the AdS$_3$ vacuum: 
\be
\label{linear_scalar}
\nabla^2 \delta W_1 =  2 e^{-W_1} [ 2 (e^{-W_2} + e^{-W_3})-e^{-W_2-W_3} ]  \delta W_1 + 4 e^{-W_1} [e^{-W_2} \delta W_2 + e^{-W_3}  \delta W_3 ].  
\ee
where $e^{W_I}$ correspond to the vacuum values. We have omitted terms cyclic in indices. 

\section{Non-SUSY spacelike warped AdS$_3$}
\setcounter{equation}{0}
\label{sec:spacelike}
In the body of this work, we have focussed on supersymmetric solutions, noting in section \ref{sec:classify} that supersymmetry has a preference for timelike warped AdS$_3$ - alternatively G\"{o}del - and warped dS$_3$ solutions. In this appendix, we relax supersymmetry in order to investigate whether the 3D U(1)$^3$ gauged supergravity permits spacelike warped AdS$_3$ solutions, which are topologically S$^1\times$AdS$_2$.  In the absence of supersymmetry, (\ref{susy_cond}) is not satisfied and the curvature of the Riemann surface, $\kappa$, becomes a free parameter. 

We begin our study by choosing the following metric, 
\be
\dd s^2_3 = \frac{\ell_1^2}{4} ( - \cosh^2 \rho \,  \dd \tau^2 + \dd \rho^2 ) + \frac{\ell_2^2}{4} ( \dd \varphi  + \sinh \rho \, \dd \tau)^2, 
\ee
where we recover AdS$_3$ with unit radius once we set $\ell_1 = \ell_2 =1$. One can next determine the Ricci tensor in orthonormal frame, 
\be
R_{00} = \frac{4}{\ell_1^2}- 2 \frac{ \ell_2^2}{\ell_1^4} = - R_{11}, \quad R_{22} = - 2 \frac{ \ell_2^2}{\ell_1^4}, 
\ee
where we have introduced the frame, $e^{0} = (\ell_1/ {2}) \cosh \rho \, \dd \tau, e^1 = ({\ell_1}/{2}) \dd \tau$ and  $e^{2} = (\ell_2/{2}) ( \dd \varphi + \sinh \rho \, \dd \tau)$. Once again, we note that when $\ell_1 = \ell_2 = 1$, we get the expected form for the Ricci tensor of AdS$_3$, namely $R_{\mu \nu} = -2 \eta_{\mu \nu}$. 

To support this geometry we now need to stipulate the ansatz for the scalar and gauge fields. Recalling the outcome of the supersymmetry classification of section \ref{sec:classify}, it is appropriate to consider constant scalars $W_I$ and field strengths $G^I$  threading AdS$_2$. To this end, we introduce constants $\beta_I$, 
\be
G^{I} = \beta_{I} \vol (\textrm{AdS}_2). 
\ee 
The equations of motion for $B^I$ can then be recast in the form of a homogeneous system of linear equations 
\be
\label{flux_EOM}
\biggl( \begin{array}{ccc} 2 \, e^{2 W_1} x & a_3 & a_2 \\ a_3 & 2 \, e^{2 W_2} x & a_1 \\ a_2 & a_1 & 2 \, e^{2 W_3} x \end{array}  \biggr) \biggl( \begin{array}{c} \beta_1 \\ \beta_2 \\ \beta_3 \end{array} \biggr) = 0, 
\ee
where we have redefined the ratio, $x = \ell_2/ \ell_1^2$. For a non-trivial solution ($\beta_I \neq 0$) to exist, we then require that the matrix be singular with zero determinant. 

We are then in a position to solve for $\beta_1$ and $\beta_2$ in terms of $\beta_3$ through Gaussian elimination. The end result is 
\be
\label{null_space}
\beta_1 = \frac{(a_1 a_3 - 2 e^{2 W_2} a_2 x)}{(4 e^{2 W_1 + 2 W_2} x^2 -a_3^2)} \beta_3, \quad \beta_2 = \frac{(a_2 a_3 - 2 e^{2 W_1} a_1  x)}{(4 e^{2 W_1 + 2 W_2} x^2 -a_3^2)} \beta_3, 
\ee
where the scalars $W_I$ are subject to the constraint: 
\be
\label{det}
4 \, e^{2 (W_1 + W_2 + W_3)} \, x^3 -  (e^{2 W_1} a_1^2 + e^{2 W_2} a_2^2 + e^{2 W_3} a_3^2) \, x +  a_1 a_2 a_3 =0,  
\ee
which is required to ensure the null space is non-trivial. A similar condition obviously holds for the supersymmetric case presented in the body of the paper. Note to perform these manipulations we have assumed the denominator does not vanish, i. e. $ x \neq \pm (a_3/2) e^{-W_1-W_2}$. To recapitulate, through (\ref{null_space}) and (\ref{det}), we have solved the EOMs for the gauge fields.  We now turn our attention to the Einstein equations.

With the earlier expressions for the Ricci tensor, the Einstein equations become 
\bea
 \label{Einstein1} \frac{4}{\ell_1^2} - 2 x^2 &=& V, \\
\label{Einstein2} - 2 x^2 &=& - \left[ V + \frac{1}{2} \sum_{I=1}^3 \partial_{W_I} V \right], 
\eea
where we have eliminated the field strengths using the scalar EOM in (\ref{Einstein2}). In contrast, they drop out completely from (\ref{Einstein1}). By combining the last equation with (\ref{det}) and the explicit expression for the potential $V$ (\ref{Einsteinact}), one can infer that $x$ must be of the form
\be
\label{x}
x =  \frac{a_1 a_2 a_3}{2 \kappa} e^{-W_1 -W_2 -W_3}.  
\ee
We remark that this is true only for generic $a_I \neq 0$, further implying that $\kappa \neq 0$, i. e. we cannot consider compactifications on a torus from 5D. The analysis with vanishing $a_I$, although more straightforward since one can easily solve (\ref{det}), one quickly finds $\kappa =a_I = \beta_I = 0$ from consistency with the EOMs. 

For the moment, we normalise $\kappa = \pm 1$, so that $\kappa^2 = 1$. Without loss of generality one can always do this since $W_I$ include a contribution from the breathing mode of the Riemann surface (\ref{scalars3D5D}). Through (\ref{x}) we have reconciled (\ref{Einstein2}) and (\ref{det}), so we have a single condition: 
\be
\label{equation1} (a_1 a_2 a_3)^2 = \left[ e^{2 W_1} a_1^2 + e^{2 W_2} a_2^2 + e^{2 W_3} a_3^2 - 2 \kappa \, e^{W_1 + W_2 + W_3} \right]. 
\ee
The remaining Einstein equation determines $\ell_1$ in terms of the scalars $W_I$, 
\be
\label{L1}
\ell_1^2 = \frac{4 e^{W_1 + W_2 + W_3}}{(4 e^{W_1} + 4 e^{W_2} + 4 e^{W_3} + \kappa)}. 
\ee
We finally must solve the scalar EOMs, an exercise that results in the following equation for $\beta_3$, 
\be
\label{beta3}
\beta_3^2 =   \frac{\left[ 2 e^{W_1 + W_2 + W_3} (2 e^{W_1} +2 e^{W_2} +\kappa)- e^{2 W_1} a_1^2 - e^{2 W_2} a_2^2  \right] }{ e^{2 W_3} \left[ 4 e^{W_1} + 4 e^{W_2} + 4 e^{W_3} + \kappa \right]^{2}}
\ee
and two further constraints on $W_I$: 
\bea
\label{equation2} a_1^2 \, e^{2 W_1} (1- \kappa \,  a_2^2 \, e^{W_2 -W_1 -W_3})^2 \partial_{W_3} V - a_3^2 \, e^{2 W_3} (1- a_1^2 a_2^2 \, e^{-2 W_3} )^2 \partial_{W_1} V &=& 0, \\
\label{equation3} a_2^2 \, e^{2 W_2} (1- \kappa \,  a_1^2 \, e^{W_1 -W_2 -W_3})^2 \partial_{W_3} V - a_3^2 \, e^{2 W_3} (1- a_1^2 a_2^2 \, e^{-2 W_3} )^2 \partial_{W_2} V &=& 0. 
\eea

In principle, one can now solve (\ref{equation1}), (\ref{equation2}) and (\ref{equation3}) for (real) $W_I$ in terms of $a_I$, before inserting expressions into (\ref{beta3}), (\ref{L1}), (\ref{x}) and (\ref{null_space}) to determine the explicit solution.  

To demonstrate that this is possible, subject to the quantisation condition for a well-defined geometry, $2 a_I (\frak{g}-1) \in \mathbb{Z}$, we truncate the 3D theory by setting $W_2 = W_1$, $\beta_2 = \beta_1$ and adopt the following parameter choice \footnote{Note since $\kappa =-1$, we can quotient the Riemann surface to increase the genus, thereby satisfying the quantisation condition.}
\be
a_1 = a_2 = \frac{3}{2} = - a_3, \quad \kappa =-1. 
\ee
Systemically solving the above equations, one can determine the explicit solution: 
\bea
e^{W_1} = e^{W_3} = \frac{9}{8}, \quad \beta_3 = \frac{3}{25} \sqrt{\frac{3}{2}}, \quad \beta_1 = - \beta_3,  \quad
\ell_1 =  \frac{27}{40}, \quad \ell_2 = \frac{27}{50}. 
\eea

We observe that the U(1) fibre is squashed, $\ell_2 < \ell_1$. As a consequence, the Killing vector $\partial_{\tau}$ is globally timelike \cite{Bengtsson:2005zj}. It is interesting to find  solutions with $\ell_1 < \ell_2$, where the Killing vector becomes spacelike at large $\rho$ and identifications give rise to black hole solutions with no CTCs outside the horizon \cite{Anninos:2008fx}. Although the above equations are difficult to solve for general $W_I$ and $a_I$, if one considers the truncation $\beta_2 = \beta_1, a_2 = a_1$ and $W_2 = W_1$, it is possible to show that $\beta_3^2 \geq 0$ precisely in the range where $\ell_1 \geq \ell_2$. When the inequalities are saturated, this is consistent with our expectations that $\ell_1 = \ell_2$ with $\beta_I = 0$. 
Within this truncation, this precludes spacelike warped AdS$_3$ solutions where the fibre is stretched. 

The above analysis involves the generic case. However, we can solve the EOMs for the field strengths by increasing the dimension of the null space. To this end, we can choose 
\be
x = \frac{a_1}{2} e^{-W_2 -W_3}, \quad a_1 e^{W_1} = a_2 e^{W_2} = a_3 e^{W_3}, 
\ee
with $a_I \neq 0$, so that there is only one relation between the $\beta_{I}$, 
\be
\frac{\beta_1}{a_1} + \frac{\beta_2}{a_2} + \frac{\beta_3}{a_3} = 0. 
\ee
We next solve the Einstein equations
\be
e^{W_1} = \frac{a_2 a_3}{\kappa}, \quad \frac{4}{\ell_1^2} = \frac{(4 a_1 a_2 + 4 a_2 a_3 + 4 a_3 a_1 +1)}{a_1^2 a_2^2 a_3^2}. 
\ee
Without loss of generality, we can take $\kappa = \pm 1$ provided we orchestrate the $a_I$ signs so that $W_I$ remain real. We finally solve the scalar EOMs, presenting us with 
\be
\beta_1 = \pm \sqrt{\frac{4 a_1^3 (a_2 + a_3)}{(4 a_1 a_2 + 4 a_2 a_3 + 4 a_3 a_1 +1)^2}}. 
\ee
with similar expressions for $\beta_2, \beta_3$. One is just left to impose the relation between the $\beta_I$. The ratio may be determined from the expression for $x$, 
\be
\frac{\ell_2}{\ell_1} =  \frac{1}{ \sqrt{(4 a_1 a_2 + 4 a_2 a_3 + 4 a_3 a_1 +1)}}. 
\ee
From the requirement that $W_I$ be real, we recognise that $a_I$ should all have the same sign, meaning that once again the U(1) fibre is squashed. 

Finally, we try one more throw of the dice to find a solution with $\ell_2 > \ell_1$; we consider the case where one of the $\beta_I$ vanish, since if two vanish, we are quickly led to a  trivial solution, $\beta_I = 0$. Choosing $\beta_3 = 0$, we can solve (\ref{flux_EOM}) by setting 
\be
x = \frac{a_3}{2 e^{W_1 + W_2}}, \quad a_1 e^{W_1} = a_2 e^{W_2}, \quad \beta_1 = - \frac{e^{W_2}}{e^{W_1}} \beta_2. 
\ee
The Einstein equations can then be solved through (\ref{L1}) and $e^{W_3-W_1 +W_2} = a_1^2$ with normalised curvature, $\kappa =1$. With the above conditions, we find it is not possible to impose $\partial_{W_3} V =0 = \partial_{W_1} V - \partial_{W_2} V $ as required by the scalar EOMs. 

This then completes our study of spacelike warped AdS$_3$ solutions to 3D U(1)$^3$ gauged supergravity. We have found spacelike warped AdS$_3$ geometries where the fibre is squashed, but not stretched. Given that our 3D solutions correspond to 5D solutions to U(1)$^3$ gauged supergravity of the form $M_3 \times \Sigma_{\frak{g}}$, the analysis also holds for the 5D spacetimes of the same form.

\end{document}